\documentclass[aps,preprint,nofootinbib]{revtex4}
\usepackage{graphicx}
\usepackage{amssymb}
\usepackage{amsmath}
\usepackage{epstopdf}
\usepackage{ulem}

\usepackage[greek,english]{babel}
\usepackage[iso-8859-7]{inputenc}

\usepackage{color}

\def\ba{\begin{eqnarray}}
\def\ea{\end{eqnarray}}
\def\be{\begin{equation}}
\def\ee{\end{equation}}

\def\gtorder{\mathrel{\raise.3ex\hbox{$>$}\mkern-14mu
             \lower0.6ex\hbox{$\sim$}}}
\def\ltorder{\mathrel{\raise.3ex\hbox{$<$}\mkern-14mu
             \lower0.6ex\hbox{$\sim$}}}

\def\dalemb#1#2{{\vbox{\hrule height.#2pt
  \hbox{\vrule width.#2pt height#1pt \kern#1pt \vrule width.#2pt}
    \hrule height.#2pt}}}

\newcommand{\planck}{{\sc Planck}} 
\newcommand{\designer}{{\sc Designer}}

\begin{document}
\rightline{T$\Theta\Delta$}


\title{Real space estimator for the weak lensing convergence from the CMB
}
\author{C.S.~Carvalho\footnote{Corresponding author: carvalho.c@gmail.com}
and K.~Moodley 
}
\affiliation{
Astrophysics and Cosmology Research Unit, School of Mathematical Sciences, 
University of KwaZulu-Natal, 4000 Durban, South Africa}

\begin{abstract}

We propose an estimator defined in real space for the reconstruction of the weak lensing potential due to the intervening large scale structure
from high resolution maps of the cosmic microwave background. This estimator was motivated 
as an alternative to the quadratic estimator in harmonic space
to surpass the difficulties 
of the analysis of maps containing galactic cuts and point source excisions. 
Using maps synthesised by pixel remapping, we implement the estimator for 
two experiments, namely one in the absence and one in the presence of detector noise,
and compare the reconstruction of the convergence field with that obtained with the quadratic estimator defined in harmonic space. We find good agreement between the input and the reconstructed power spectra using the proposed real space estimator. We discuss interesting features of the real space estimator and future extensions of this work.














\end{abstract}
\date{\today}
\maketitle

\section{Introduction}


Observations of the cosmic microwave background (CMB) are a powerful probe of the physics of the early universe and a robust discriminant of cosmological models. 
Since the CMB photons propagate more or less freely after recombination, what we observe on large scales are mainly the perturbations on the last-scattering surface. 
The CMB contains also an imprint of the photon quadrupole which developed 
during the recombination epoch and generated linear polarisation through the mechanism of Thomson scattering.

However, the CMB should also contain signatures of processes that occurred between the last-scattering surface and the present. These processes will be manifest as non-linear signals which become important on small scales. 
The growing improvement in resolution and sensitivity of CMB experiments allows us to resolve ever so smaller scales and motivates the development of estimators 
and 
robust implementations for the extraction of the different signals.
Among such processes is weak gravitational lensing, which consists of 
the deflection of CMB photons 
by the mass of matter clustered along the line of sight,
while leaving the surface brightness unaffected. 
This effect distorts the temperature-temperature (TT) correlation power spectrum by a few arcminutes but coherently over degree scales \cite{seljak}, 
which means that lensing becomes an important effect for $\ell\ge 3000.$ 
Weak lensing also mixes and distorts the EE and BB polarisation power spectra
\cite{zaldarriaga&seljak}. 
Finally, weak lensing also
correlates with other secondary effects, such as the integrated Sachs-Wolfe (ISW) effect 
and the Sunyaev-Zel'dovich (SZ) effect, 
thus inducing non-Gaussianities which are manifest in higher-order point correlation functions \cite{zaldarriaga,cooray}.
 
The relevance of the CMB lensing reconstruction for cosmology is threefold. 
First, by converting a fraction of the dominant E-mode polarisation due to density perturbations into a B-mode polarisation, CMB lensing introduces a contaminant in the measurement of the primordial B-mode polarisation and consequently of the inflation energy scale 
\cite{gravitational_waves}.
Second, CMB lensing is also a powerful cosmological probe of the matter distribution integrated from the last scattering surface at redshift $z\sim 1000$ to the present time.
Third, by changing the power spectra of the perturbations and inducing non-Gaussianities, CMB lensing affects cosmological parameter estimation.
Delensing the CMB will thus allow one to recover the primordial B-mode, probe the full-sky large-scale structure distribution
with a maximum efficiency at $z\approx 3$ \cite{lewis&challinor}, and obtain unbiased parameter constraints from CMB measurements \cite{lewis}.


Effort has been devoted to developing an optimal reconstruction of the lensing potential in harmonic space, which implicitly assumes full-sky coverage without galactic cuts, bad pixels due to excision of point sources or 
nonuniform weighting for uneven sky coverage \cite{okamoto&hu}.
In this idealized context, it has been shown how to construct an optimal quadratic estimator for a reconstruction based on the temperature anisotropy alone \cite{hu} (see \cite{hu&okamoto} for an extension of the formalism to include the polarisation). The improvement gained from using a maximum likelihood, and thus more optimal, estimator for the reconstruction from the temperature is marginal \cite{hirata&seljak2002} (although this might not necessarily be the case for experiments sensitive to the B-mode polarisation \cite{hirata&seljak2003}). This is because the distortion due to lensing is small compared to the intrinsic cosmic variance or to the experimental noise.


Our interest lies in considering a slightly less optimal estimator modified to have a finite range in real space. The estimator is based on a convolution of the square of the lensed temperature map with a local kernel, as has been suggested previously \cite{bucher}.
An estimator with kernel in real space acts locally, as opposed to the conventional estimator with kernel in harmonic space which acts over the entire celestial sphere even though it has finite support in harmonic space. 
The extent in real space will be the inverse of that in harmonic space, thus finite and determined by the scale of the root mean square of the lensing 
convergence, with the contribution from sources beyond the kernel being negligible. An estimator with these properties can easily be adapted to include cuts, excisions of pixels and nonuniform sky coverage
for a local extraction of the lensing 
convergence. In particular, 
a nonuniform sky coverage can be accounted for by a corresponding weighting in the real space convolution.

The lensing distortion of the CMB anisotropy on the surface of last scattering can be described in three ways: by the lensing potential $\psi,$ by the deflection vector 
$\boldsymbol{\alpha}=\nabla \psi$ or by the convergence tensor $\boldsymbol{\kappa}=-\nabla\nabla\psi/2.$ The descriptions of 
$\psi$ and $\boldsymbol{\alpha}$ suffer from an ambiguity upon translation, since a patch of the sky and its translation have the same likelihood on account of isotropy. In contrast, the description of the convergence, which is a gradient of the deflection vector field, is locally well defined. 
Here we 
implement the real space estimator to reconstruct the convergence 
and compare the reconstruction to that obtained with the harmonic space estimator.

The convergence tensor can be decomposed as a sum of an isotropic (diagonal) tensor and an anisotropic (traceless) tensor as follows
\ba
\boldsymbol{\kappa}
=\left(
\begin{array}{cc}
\kappa_{0}+\kappa_{+} & \kappa_{\times}\\
\kappa_{\times} & \kappa_{0}-\kappa_{+}
\end{array}
\right).
\ea
For a small patch of sky, we can use the flat-sky approximation and define a Cartesian coordinate system with basis vectors $\{\boldsymbol{e}_{x}(r),\boldsymbol{e}_{y}(r)\}$ and metric $g={\rm diag}(1,1).$ 
In particular, the convergence $\kappa_{0}=-(\partial_{x}^2+\partial_{y}^2)\psi/2$ magnifies a feature on the last-scattering surface, the shear $\kappa_{+}=-(\partial_{x}^2-\partial_{y}^2)\psi/2$ stretches it along the $x$--axis while compressing it along the $y$--axis, and analogously the shear $\kappa_{\times}=-\partial_{x}\partial_{y}\psi$ stretches it along the $y=x$ axis and compresses it along the $y=-x$ axis.
Here we focus on the dilation effect
captured by the isotropic component $\kappa_{0},$
and will treat the shear components in a forthcoming study.

The convergence can be further decomposed into two contributions, namely the convergence of the unlensed temperature anisotropies, to a high degree isotropic, and the convergence of the lensing field, by nature random. 
In order to remove the zero mode of the unlensed CMB, we average the convergence  on each local patch over the sky region surveyed and subtract this average value from the estimated convergence map. 
Higher multipoles of the unlensed CMB will still be present in the estimated convergence map, which give rise to a noise bias. This bias can be removed by cross-correlating the reconstructed map with maps of the large scale structure derived from independent methods, e.g. a galaxy redshift survey. It can also be removed when estimating the convergence power spectrum by combining, in harmonic space, different triangle configurations with a common side.
The formulation of an estimator for the lensing convergence given the lensed image still requires some prior knowledge of the statistics of the
underlying temperature field, in particular 
the power spectrum of the unlensed sky which is assumed to be Gaussian.

This is the first of a series of studies where we explore the properties of the proposed estimator and test its implementation.
The manuscript is organized as follows.
In Section~\ref{sec:weak_lensing}
we 
describe the weak lensing CMB of photons in the Born approximation. 
In Section~\ref{sec:real_estimator} 
we derive the proposed estimator defined in real space, whose kernel in harmonic space is described in Appendix~\ref{app:harmonic_estimator_description}.
In Section~\ref{sec:reconstruction} we proceed to reconstruct the convergence power spectrum by implementing the real space estimator on lensed maps synthesised as detailed in Appendix~\ref{app:map_synthesis}. For comparison, we also implement the harmonic space estimator 
as detailed in Appendix~\ref{app:harmonic_estimator_implementation}.
We discuss the results obtained from the two estimators in Section~\ref{sec:discussion} and conclude  in Section~\ref{sec:summary} by discussing further developments of the real space estimator in follow-up studies.

\section{Weak lensing of the CMB}
\label{sec:weak_lensing}

Weak lensing of the 
CMB consists of the deflection of CMB photons from the original propagation direction $\boldsymbol{\theta}$ on the last scattering surface (the source plane) to an observed direction $\tilde{\boldsymbol{\theta}}$ on the sky today (the image plane).
This deflection amounts to remapping the unlensed temperature anisotropies $T$ to the lensed ones $\tilde T$ according to 
\ba
\tilde T(\boldsymbol{\theta})
=T(\tilde{\boldsymbol{\theta}}).
\ea
The deflection angle 
${\boldsymbol{\alpha}}= \tilde{\boldsymbol{\theta}}-\boldsymbol{\theta}$
is given by ${\boldsymbol{\alpha}}=\nabla\psi,$
where the lensing potential $\psi$ is given by the gravitational potential projected on the image plane and integrated along the line of sight, and $\nabla$ is the covariant derivative on the image plane.
The lensing potential measures the cumulative effect of deflectors along the line of sight which
lens the perturbations at the last scattering surface
and thus lensed are imprinted on the CMB.
Consider the line element of a homogeneous and isotropic (1+3)--dimensional spacetime subject to scalar perturbations expanded up to linear order
\ba
ds^2=a^2(\eta)\left[ -(1-2\Psi)d\eta^2 
+(1+2\Psi)\left( dr^2+f^2_{k}(r)~d\Omega_{k}^2\right)\right],
\ea
where the functional form of $f_{k}$ depends on the spatial curvature $k$ of the two-dimensional surface $\Omega_{k}$ 
as follows
\ba
f_{k}(r)=\left\{
\begin{array}
{rr}
\sin[r] : &{\rm for}~k=+1,\\
r : &{\rm for}~k=0,\\
\sinh[r] : & {\rm for}~ k=-1.
\end{array}
\right.
\ea
Integrating the equation for a null geodesic in an arbitrarily curved spacetime, 
we find for the lensing potential $\psi$ generated by the gravitational potential $\Psi$ that
\cite{lewis&challinor}
\ba
\psi(\boldsymbol{\theta})
=-2\int ^{r_{LS}}_{0} dr
{f_{k}(r_{LS}-r)\over {f_{k}(r_{LS})~f_{k}(r)}}
\Psi(r\boldsymbol{\theta},r),
\ea
where $r$ is the conformal distance to the lens plane, $r_{LS}$ is the conformal distance to the last scattering surface and $\boldsymbol{\theta}$ is the two-dimensional position vector on the image plane.
This expression was derived by integrating the photon geodesics from the source along the unperturbed path 
(the Born approximation). 
The gravitational lensing by $\psi$ 
deflects the photons on the image plane
by $\nabla \psi,$ thus
remapping the temperature anisotropies according to
\ba
\tilde T(\boldsymbol{\theta})
=T(\boldsymbol{\theta}+\nabla \psi)
=T(\boldsymbol{\theta})
+\nabla\psi \cdot \nabla T(\boldsymbol{\theta})
+O[(\nabla\psi)^2].
\ea
Since the Born approximation assumes that $\nabla\psi $ is constant between $\boldsymbol{\theta}$ and $\tilde{\boldsymbol{\theta}}$, it is only valid for deflections small compared to the scale of the lensing perturbations, i.e. deflections up to a few arcminutes. 

Considering the full sky described as a two-sphere centred at the observer, then the temperature anisotropies can be expanded in spherical harmonics as follows
\ba
T(\boldsymbol{\theta})
=\sum_{\ell=0}^{\infty}\sum_{m=-\ell}^{\ell}
a_{\ell m}~Y_{\ell m}(\Omega_{k}),
\ea
where $\boldsymbol{\theta}$ 
denotes the position on the two-sphere.
In the flat-sky approximation, the CMB temperature anisotropy is instead expanded in terms of plane waves as follows
\ba
T(\boldsymbol{\theta})
={1\over {\sqrt{4\pi}}}\sum_{\ell=0}^{\infty}
a_{\ell}\exp[i\boldsymbol{\ell}\cdot\boldsymbol{\theta}]
\ea
where $\boldsymbol{\theta}=(\theta_{x},\theta_{y}).$ 
To obtain the same asymptotic density of states as on the full-sky description, we compactify the sphere to a square having sides of length $L=\sqrt{4\pi}$ so that the discretized variable $\boldsymbol{\theta}/L$ is replaced by the continuous wave number $\boldsymbol{\ell}/(2\pi),$ yielding the Fourier mode expansion
\ba
T(\boldsymbol{\theta})
=\int {d^2\boldsymbol{\ell}\over (2\pi)^2}~
\exp[i\boldsymbol{\ell}\cdot\boldsymbol{\theta}]~T(\boldsymbol{\ell}).
\ea
This is a valid approximation for sufficiently small angular scales 
where we can replace the expansion in spherical harmonics by an expansion in plane waves. 
At the largest angular scales, however, the curvature and connectivity properties of the celestial sphere are no longer negligible and the approximation is no longer valid.
Likewise expanding the lensing potential in plane waves
so that 
$(\nabla \psi)(\boldsymbol{\ell})
=i\ell\psi(\boldsymbol{\ell}),
$
we can write the linear order correction of the lensed temperature anisotropy in harmonic space as the convolution of the lensing potential with the unlensed temperature anisotropy 
\ba
\tilde T(\boldsymbol{\ell})
=T(\boldsymbol{\ell})
-\int d^2\boldsymbol{\ell}^{\prime}~
\boldsymbol{\ell}^{\prime}\cdot (\boldsymbol{\ell}-\boldsymbol{\ell}^{\prime})~
\psi(\boldsymbol{\ell}^{\prime})~
T(\boldsymbol{\ell}-\boldsymbol{\ell}^{\prime})
+O[\psi^2(\boldsymbol{\ell})].
\ea

The lensing effects are apparent in the temperature power spectrum.
Defining the temperature power spectrum by 
$\left<T(\boldsymbol{\ell})~T(\boldsymbol{\ell}^{\prime})\right>
=(2\pi)^2~
\delta(\boldsymbol{\ell}+\boldsymbol{\ell}^{\prime})~
C_{\ell} ,$
we find that the power spectrum of the lensed temperature anisotropy is related to that of the unlensed as follows
\ba
\left< \tilde T(\boldsymbol{\ell}^{\prime})~
\tilde T(\boldsymbol{\ell}-\boldsymbol{\ell}^{\prime})
\right>
=(2\pi)^2~
\delta(\boldsymbol{\ell})~C_{\ell^{\prime}}
+(2\pi)^2
\left[ \boldsymbol{\ell}\cdot\boldsymbol{\ell}^{\prime}~
C_{\ell^{\prime}}
+\boldsymbol{\ell}\cdot(\boldsymbol{\ell}-\boldsymbol{\ell}^{\prime})~
C_{\vert \boldsymbol{\ell}-\boldsymbol{\ell}^{\prime}\vert}\right]
\psi(\boldsymbol{\ell}).
\label{eqn:c_tt}
\ea
Since the convolution is made of gradient terms, it results in the transfer of power from large to small scales, with the random deflections simultaneously smearing out the sharp features of the unlensed spectrum and thus leading to a suppression of the acoustic oscillations \cite{seljak}.
The first term in Eqn.~(\ref{eqn:c_tt}) describes a statistically isotropic ensemble where modes with different $\boldsymbol{\ell}$ are uncorrelated.
Anisotropies, and in particular those generated by the lensing potential, introduce correlations among different $\boldsymbol{\ell}$ modes.
What we will be using to reconstruct the lensing field are these off-diagonal correlations.

\section{Estimator of the lensing convergence in real space}
\label{sec:real_estimator}

\subsection{Derivation of the real space estimator}

In harmonic space
the estimator of the lensing potential $\hat \psi(\boldsymbol{\ell})$ is expressed as the convolution of the square of the lensed map $\tilde T(\boldsymbol{\ell})$
by a weight function 
$Q^{\psi}(\boldsymbol{\ell},\boldsymbol{\ell}^\prime)$ as follows 
\ba
\hat \psi(\boldsymbol{\ell})
=\int {d^2\boldsymbol{\ell}^{\prime}\over (2\pi)^2}~
\tilde T(\boldsymbol{\ell}^\prime)~\tilde T(\boldsymbol{\ell}-\boldsymbol{\ell}^\prime)~
 Q^{\psi}(\boldsymbol{\ell},\boldsymbol{\ell}^\prime).
\label{eqn:psi_l}
\ea
The optimal estimator is found for 
the weight function which minimizes the leading order variance, which is given by
\ba
Q^{\psi}(\boldsymbol{\ell},\boldsymbol{\ell}^{\prime})
=\mathcal{N}_{\ell}~{1\over 2}
{ {\boldsymbol{\ell}\cdot\boldsymbol{\ell}^{\prime}C_{\ell^{\prime}}
+\boldsymbol{\ell}\cdot(\boldsymbol{\ell}-\boldsymbol{\ell}^{\prime})
 C_{\vert\boldsymbol{\ell}-\boldsymbol{\ell}^{\prime}\vert}
}\over 
{ [\tilde C_{\ell^{\prime}}+N_{\ell^{\prime}}]
[\tilde C_{\vert\boldsymbol{\ell}-\boldsymbol{\ell}^{\prime}\vert}
+N_{\vert\boldsymbol{\ell}-\boldsymbol{\ell}^{\prime}\vert}]
} }
\label{eqn:q_ell}
\ea
where $N_{\ell}$ is the detector noise power spectrum given in Eqn.~(\ref{eqn:n_ell})
and $\mathcal{N}_{\ell}$ is the variance of the estimator given in Eqn.~(\ref{eqn:norm_factor})
(see Appendix \ref{app:harmonic_estimator_description} for a derivation). 
The relation between the convergence and the lensing potential in harmonic space  translates into the same relation for the corresponding estimators. 
Hence the weight function for the convergence estimator will be
\ba 
Q(\boldsymbol{\ell},\boldsymbol{\ell}^{\prime})
=
\ell^2Q^{\psi}(\boldsymbol{\ell},\boldsymbol{\ell}^{\prime}).
\ea

Naively we would write the corresponding estimator in real space as
\ba
\hat \kappa_{0} (\boldsymbol{\theta})
&=&
\int {d^2\boldsymbol{\ell}\over (2\pi)^2}~
 \exp[i\boldsymbol{\ell}\cdot\boldsymbol{\theta}]
  \int {d^2\boldsymbol{\ell}^{\prime}\over (2\pi)^2}~
  \tilde T(\boldsymbol{\ell}^{\prime})~
   \tilde T(\boldsymbol{\ell}-\boldsymbol{\ell}^{\prime})~
    Q(\boldsymbol{\ell},\boldsymbol{\ell}^{\prime}) \cr
&\equiv&
\int d^2\boldsymbol{\theta}^{\prime}~
\tilde T(\boldsymbol{\theta}^{\prime})
 \int d^2\boldsymbol{\theta}^{\prime\prime}~
 \tilde T(\boldsymbol{\theta}^{\prime\prime})~
  Q(\boldsymbol{\theta},\boldsymbol{\theta}^{\prime},
  \boldsymbol{\theta}^{\prime\prime}),
\label{eqn:kappa_q}
\ea
where we define the corresponding weight function in real space 
by 
\ba
Q(\boldsymbol{\theta},\boldsymbol{\theta}^{\prime},
\boldsymbol{\theta}^{\prime\prime})
= \int {d^2\boldsymbol{\ell}\over (2\pi)^2}~
 \exp[i\boldsymbol{\ell}\cdot\boldsymbol{\theta}]
   \int {d^2\boldsymbol{\ell}^{\prime}\over (2\pi)^2}~
     \exp[-i\boldsymbol{\ell}^{\prime}\cdot\boldsymbol{\theta}^{\prime}]
     \exp[-i(\boldsymbol{\ell}-\boldsymbol{\ell}^{\prime})\cdot
      \boldsymbol{\theta}^{\prime\prime}]~
      Q(\boldsymbol{\ell},\boldsymbol{\ell}^{\prime}).
\label{eqn:q_theta}
\ea
In general, $Q(\boldsymbol{\theta},\boldsymbol{\theta}^{\prime},\boldsymbol{\theta}^{\prime\prime})$ would be a function of six variables, namely the three lengths and the three corresponding angles of the three vectors $\boldsymbol{\theta},$ $\boldsymbol{\theta}^{\prime}$ and 
$\boldsymbol{\theta}^{\prime\prime}.$ 
However, the functional dependence of $Q(\boldsymbol{\theta},\boldsymbol{\theta}^{\prime},
\boldsymbol{\theta}^{\prime\prime})$ is derived from $Q(\boldsymbol{\ell},\boldsymbol{\ell}^{\prime})$
which depends on three variables, namely the lengths of $\boldsymbol{\ell}$ and $\boldsymbol{\ell}^{\prime},$ and the angle between them  $\xi_{\boldsymbol{\ell}}
=\phi_{\boldsymbol{\ell}}-\phi_{\boldsymbol{\ell}^{\prime}},$ due to the fact that the vectors in harmonic space form a triangle. 
This reduces  $Q(\boldsymbol{\theta},\boldsymbol{\theta}^{\prime},\boldsymbol{\theta}^{\prime\prime})$ to a function of three variables, namely the length of two of the vectors $\boldsymbol{\theta}$ and $\boldsymbol{\theta}^{\prime},$ and the angle between them  $\xi_{\boldsymbol{\theta}}
=\phi_{\boldsymbol{\theta}}-\phi_{\boldsymbol{\theta}^{\prime}}.$
We can expand the kernel in terms of eigenfunctions that factorize the radial and the angular dependence as follows 
\ba
Q(\boldsymbol{\ell},\boldsymbol{\ell}^{\prime})
=\sum_{m=-\infty}^{+\infty}\exp[im\xi_{\boldsymbol{\ell}}]~
Q_{m}(\ell,\ell^{\prime}),
\ea
so that
\ba
Q_{m}(\ell,\ell^{\prime})
={1\over 2\pi}\int d\xi_{\boldsymbol{\ell}}~\exp[-im\xi_{\boldsymbol{\ell}}]~
Q(\boldsymbol{\ell},\boldsymbol{\ell}^{\prime}).
\ea
Let $\phi_{\boldsymbol{\ell}}$ and $\phi_{\boldsymbol{\ell}^{\prime}}$ be the angles that $\boldsymbol{\ell}$ and $\boldsymbol{\ell}^{\prime}$ respectively make with the $\ell_x$--axis, and likewise
$\phi_{\boldsymbol{\theta}}$ and $\phi_{\boldsymbol{\theta}^{\prime}}$ the angles that $\boldsymbol{\theta}$ and $\boldsymbol{\theta}^{\prime}$ respectively make with the $\theta_x$--axis.
By construction 
\ba
\boldsymbol{\ell}^{\prime}\cdot\boldsymbol{\theta}^{\prime}
&=&\ell^{\prime}\theta^{\prime}
 \cos[\phi_{\boldsymbol{\theta}^{\prime}}-\phi_{\boldsymbol{\ell}^{\prime}}],
\\
(\boldsymbol{\ell}-\boldsymbol{\ell}^{\prime})\cdot\boldsymbol{\theta}^{\prime\prime}
&=&\ell\theta^{\prime\prime}
 \cos[\phi_{\boldsymbol{\ell}}-\phi_{\boldsymbol{\theta}^{\prime\prime}}]
-\ell^{\prime}\theta^{\prime\prime}
 \cos[\phi_{\boldsymbol{\ell}^{\prime}}-\phi_{\boldsymbol{\theta}^{\prime\prime}}].
\ea
Without loss of generality, let us also consider $\boldsymbol{\theta}=0.$ Then Eqn.~(\ref{eqn:q_theta}) becomes
\ba
Q(\theta^{\prime},\theta^{\prime\prime},\xi_{\boldsymbol{\theta}})
&\equiv&
\sum_{m=-\infty}^{+\infty}\exp[im\xi_{\boldsymbol{\theta}}]~
Q_{m}(\theta^{\prime},\theta^{\prime\prime},\xi_{\boldsymbol{\theta}}),
\label{eqn:q_theta2}
\ea
where
\ba
Q_{m}(\theta^{\prime},\theta^{\prime\prime},\xi_{\boldsymbol{\theta}})
&=&{1\over {(2\pi)^2}}
 \int_{0}^{\infty} d\ell~\ell
  \int_{0}^{\infty} d\ell^{\prime}~\ell^{\prime}~
   J_{m}( \ell\theta^{\prime\prime})~
    J_{m}( \ell^{\prime}(\theta^{\prime}-\theta^{\prime\prime}
    \cos[\xi_{\boldsymbol{\theta}}]) )~
     Q_{m}(\ell,\ell^{\prime})
\ea
and $\xi_{\boldsymbol{\theta}}
=\phi_{\boldsymbol{\theta}^{\prime}}-\phi_{\boldsymbol{\theta}^{\prime\prime}}$ 
is the angle between $\boldsymbol{\theta}^{\prime}$ and 
$\boldsymbol{\theta}^{\prime\prime}.$
\footnote{Here we used the Jacobi-Anger expansion of the Bessel functions \cite{abramowich}
\ba
\exp[i\ell\theta\cos[\phi]]
=\sum_{n=-\infty}^{\infty}i^{n}J_{n}(\ell\theta)\exp[in\phi]
=J_{0}(\ell\theta)+2\sum_{n=1}^{\infty}i^{n}J_{n}(\ell\theta)\cos[n\phi]
\ea
and the orthogonality condition 
\ba
\int d\phi~\exp[i\ell\theta\cos[\phi]]\exp[in^{\prime}\phi]
=\int d\phi \sum_{n=-\infty}^{\infty}i^{n}J_{n}(\ell\theta)\exp[i(n+n^{\prime})\phi]
=(2\pi)i^{n^{\prime}}\delta_{nn^{\prime}}J_{n}(\ell\theta).\quad
\ea}
The eigenfunctions 
$Q_{m}(\theta^{\prime},\theta^{\prime\prime},\xi_{\boldsymbol{\theta}})$ depend explicitly on $\xi_{\boldsymbol{\theta}}$ which means that the factorization of the angular and radial dependence of the kernel in harmonic space failed to translate into the same factorization of the kernel in real space. The computation of the estimator in Eqn.~(\ref{eqn:kappa_q}) requires a quadruple numerical integral
\ba
\kappa_{0}(\boldsymbol{\theta}=0)
&=&
\int d^2{\boldsymbol{\theta}^{\prime}}
\int d^2{\boldsymbol{\theta}^{\prime\prime}}
 \sum_{m=-\infty}^{+\infty}\exp[im\xi_{\boldsymbol{\theta}}]~
 \tilde T({\boldsymbol{\theta}^{\prime}})~
  \tilde T({\boldsymbol{\theta}^{\prime\prime}})~
   Q_{m}(\theta^{\prime},\theta^{\prime\prime},\xi_{\boldsymbol{\theta}})
\label{eqn:kappa_q2}
\ea
which involves a number of operations proportional to the number of components of $Q_{m},$ hence cubic in the size of $Q_{m}.$ 
In order to minimize the computational costs, 
we would like instead to write the real space estimator in such a form that the radial eigenfunctions in real space depend only on the length of the vectors and not on the angle between them.

To achieve this we change to the variables $\boldsymbol{\ell}_{+}, \boldsymbol{\ell}_{-}$ such that 
$\boldsymbol{\ell}=\boldsymbol{\ell}_{+},$ 
$\boldsymbol{\ell}^{\prime}=(\boldsymbol{\ell}_{+}+\boldsymbol{\ell}_{-})/2.$
In the coordinates $(\boldsymbol{\ell}_{+},\boldsymbol{\ell}_{-})$ the kernel becomes squeezed in the $\boldsymbol{\ell}_{-}$ direction as $\boldsymbol{\ell}_{+}$ increases. 
Then Eqn.~(\ref{eqn:psi_l}) yields
\ba
\hat \kappa_{0}(\boldsymbol{\ell}_{+})
=\int {d^2\boldsymbol{\ell}_{-}\over (2\pi)^2}~
\tilde T\left( {\boldsymbol{\ell}_{+}+\boldsymbol{\ell}_{-}}\over 2\right)
 \tilde T\left( {\boldsymbol{\ell}_{+}-\boldsymbol{\ell}_{-}}\over 2\right)
 W(\boldsymbol{\ell}_{+},\boldsymbol{\ell}_{-}),
\ea
where
\ba
W(\boldsymbol{\ell}_{+},\boldsymbol{\ell}_{-})
=
Q(\boldsymbol{\ell},\boldsymbol{\ell}^{\prime})
\label{eqn:w-q}
\ea
upon change of variables. The corresponding estimator in real space becomes
\ba
\hat \kappa_{0} (\boldsymbol{\theta})
&\equiv&
\int d^2\boldsymbol{\theta}_{+}
 \int d^2\boldsymbol{\theta}_{-}~
  \tilde T(\boldsymbol{\theta}_{+}+\boldsymbol{\theta}_{-})~
   \tilde T(\boldsymbol{\theta}_{+}-\boldsymbol{\theta}_{-})~
    W(\boldsymbol{\theta},\boldsymbol{\theta}_{+},\boldsymbol{\theta}_{-}),
\label{eqn:kappa_w}
\ea
where we define the corresponding weight function in real space 
by
\ba
W(\boldsymbol{\theta},\boldsymbol{\theta}_{+},\boldsymbol{\theta}_{-}) 
&=&
\int {d^2\boldsymbol{\ell}_{+}\over (2\pi)^2}~
\exp[ i\boldsymbol{\ell}_{+}\cdot\boldsymbol{\theta}]
 \int {d^2\boldsymbol{\ell}_{-}\over (2\pi)^2}~
  \exp\left[ -i(\boldsymbol{\ell}_{+}\cdot\boldsymbol{\theta}_{+}
  +\boldsymbol{\ell}_{-}\cdot\boldsymbol{\theta}_{-})\right]
  W(\boldsymbol{\ell}_{+},\boldsymbol{\ell}_{-}). \qquad
\label{eqn:w_theta}
\ea

Again we expand the kernel $W(\boldsymbol{\ell}_{+},\boldsymbol{\ell}_{-})$ in terms of eigenfunctions that factorize the radial and the angular dependence 
such that
\ba
W_{m}(\ell_{+},\ell_{-}) 
={1\over 2\pi}\int d\chi_{\boldsymbol{\ell}}~\exp[-im\chi_{\boldsymbol{\ell}}]~
W(\boldsymbol{\ell}_{+},\boldsymbol{\ell}_{-})
\label{eqn:w_m_ell}
\ea
and  $\chi_{\boldsymbol{\ell}}$ is the angle between $\boldsymbol{\ell}_{+}$ and $\boldsymbol{\ell}_{-}.$ 
Let $\phi_{\boldsymbol{\ell}_{+}}$ and $\phi_{\boldsymbol{\ell}_{-}}$ be the angles that $\boldsymbol{\ell}_{+}$ and $\boldsymbol{\ell}_{-}$ respectively make with the $\ell_x$--axis,  and likewise $\phi_{\boldsymbol{\theta}_{+}}$ and $\phi_{\boldsymbol{\theta}_{-}}$ the angles that $\boldsymbol{\theta}_{+}$ and $\boldsymbol{\theta}_{-}$ respectively make with the $\theta_x$--axis. 
By construction
\ba
\boldsymbol{\ell}_{+}\cdot\boldsymbol{\theta}_{+}
+\boldsymbol{\ell}_{-}\cdot\boldsymbol{\theta}_{-}
=\ell_{+}\theta_{+}\cos[\phi_{\boldsymbol{\ell}_{+}}-\phi_{\boldsymbol{\theta}_{+}}]
+\ell_{-}\theta_{-}\cos[\phi_{\boldsymbol{\ell}_{-}}-\phi_{\boldsymbol{\theta}_{-}}].
\ea
For the case that $\boldsymbol{\theta}=0,$ Eqn.~(\ref{eqn:w_theta}) becomes
\ba
W(\theta_{+},\theta_{-},\chi_{\boldsymbol{\theta}}) 
&\equiv&
\sum_{m=-\infty}^{+\infty}\exp[im\chi_{\boldsymbol{\theta}}]~
W_{m}(\theta_{+},\theta_{-}),
\label{eqn:w_theta2}
\ea
where
\ba
W_{m}(\theta_{+},\theta_{-})
&=&{1\over {(2\pi)^2}}
\int_{0}^{\infty} d\ell_{+}\ell_{+}\int_{0}^{\infty} d\ell_{-}\ell_{-}~
  J_{m}(\ell_{+}\theta_{+})~J_{m}(\ell_{-}\theta_{-})~
   W_{m}(\ell_{+},\ell_{-})
\label{eqn:w_m_theta}
\ea
and $\chi_{\boldsymbol{\theta}}=\phi_{\boldsymbol{\theta}_{+}}-\phi_{\boldsymbol{\theta}_{-}}$ is the angle between 
$\boldsymbol{\theta}_{+}$ and $\boldsymbol{\theta}_{-}.$
We have now achieved the factorization of the angular and the radial dependence of the kernel in real space, so that Eqn.~(\ref{eqn:kappa_w}) can be written as 
\ba
\hat \kappa_{0}(\boldsymbol{\theta}=0)
=
\sum_{m=-\infty}^{+\infty}
\int d^2{\boldsymbol \theta}_{+}\int d^2{\boldsymbol \theta}_{-}
 \exp[im\chi_{\boldsymbol{\theta}}]~
  \tilde T({\boldsymbol \theta}_{+}+{\boldsymbol \theta}_{-})~
   \tilde T({\boldsymbol \theta}_{+}-{\boldsymbol \theta}_{-})~
    W_{m}(\theta_{+},\theta_{-}).\quad
\label{eqn:kappa_w2}
\ea
In contrast with 
Eqn.~(\ref{eqn:kappa_q2}), 
this integral involves a number of operations proportional to the square of the size of $W_{m}.$

The convergence map thus estimated contains two contributions: the convergence derived from the lensing potential and a gaussian noise derived from the unlensed CMB, $\hat \kappa_{0}=\kappa_{0}\vert_{\psi}+\kappa_{0}\vert_{\psi=0}$ (as we will see in Section~\ref{sec:discussion}, the (white) detector noise averages out for our pixel based estimator). Taking the average of the convergence map (with a fixed lensing potential) over many realisations of the CMB results in the gaussian CMB noise averaging out so that our estimator is unbiased $\left< \hat \kappa_{0}\right>=\left< \kappa_{0}\vert_{\psi}\right>.$ However, since in practice we only have access to a single realisation of the sky, our reconstructed convergence map will be noisy.
In the absence of lensing, the convergence estimator defined in Eqn.~(\ref{eqn:kappa_w2}) will have a non-vanishing average value. We therefore subtract the average value of the map reconstructed in the absence of lensing so that our convergence estimator becomes
\ba
\hat \kappa_{0}(\boldsymbol{\theta}=0)
\!&=&
\sum_{m=-\infty}^{+\infty}
\int d^2{\boldsymbol \theta}_{+}\int d^2{\boldsymbol \theta}_{-}
 \exp[im\chi_{\boldsymbol{\theta}}]\cr
&&\times
 \left[
  \tilde T({\boldsymbol \theta}_{+}+{\boldsymbol \theta}_{-})
   \tilde T({\boldsymbol \theta}_{+}-{\boldsymbol \theta}_{-})
  -\big< \tilde T({\boldsymbol \theta}_{+}+{\boldsymbol \theta}_{-})
   \tilde T({\boldsymbol \theta}_{+}-{\boldsymbol \theta}_{-})\big> _{\kappa_{0}=0}\right]\!
    W_{m}(\theta_{+},\theta_{-}),\qquad
\label{eqn:kappa_w3}
\ea
where the average is taken over many realisations of the unlensed CMB.

We are interested in estimating the power spectrum $C_\ell^{\kappa_{0}\kappa_{0}}$ from the reconstructed convergence map. In the absence of lensing there is a non-vanishing contribution to the estimator from the collapsed four point function $\left<\kappa_{0}\vert_{\psi=0}~\kappa^{\ast}_{0}\vert_{\psi=0}\right>$ of the unlensed temperature field, which biases the power spectrum of the estimated convergence. This bias is precisely the variance of the gaussian CMB noise in the reconstructed convergence map \cite{noise_bias}. To remove this bias we generate a set of unlensed temperature maps (based on the unlensed temperature power spectrum) and compute the convergence power spectrum for each of these maps. The average of these power spectra is then subtracted from the power spectrum of the estimated convergence to yield an unbiased estimate of the convergence power spectrum $\left<\hat \kappa_{0}~\hat \kappa^{\ast}_{0}\right>=
\left<\kappa_{0}\vert_{\psi}~\kappa^{\ast}_{0}\vert_{\psi}\right>.$ 


In this formulation, 
the kernel in Eqn.~(\ref{eqn:kappa_w3}) is essentially the Bessel transform of the optimal weight function in harmonic space. 
Contrary to the formulation in harmonic space, where the temperature map is transformed before being weighted by the kernel in harmonic space, here it is the kernel that is transformed before acting on the temperature map to yield the lensing potential. The kernel weights the contribution of each pair of points on the temperature map to the weak lensing. 
Since weak lensing is manifested essentially at very small scales in the temperature map, all the information relevant to lensing reconstruction lies on angular scales close to the resolution scale of the temperature map, which is set by the beam size for both the harmonic and real space estimators. In the case of the real space estimator 
there is an additional effect that limits the reconstruction on small scales, which arises from the fact that we are averaging the squared temperature map over the spatial extent of the kernel. Since the kernel changes sign, the averaging over quadratic combinations of closely separated pixels results in a loss of recovered power on angular scales smaller than the spatial extent of the kernel. This effect has been studied in more detail in Ref.~{\cite{bucher2010}}. The angular resolution of the experiment limits the spatial extent of the kernel so that an experiment with higher angular resolution or higher sensitivity will have a more local kernel.


\section{Reconstruction of the convergence map}
\label{sec:reconstruction}


We proceed to describe the implementation of this estimator to reconstruct the convergence field $\kappa_{0}.$ 
First we synthesise lensed CMB maps as described in Appendix~\ref{app:map_synthesis}.
We then implement the real space estimator to the maps
by applying the procedure described below, and produce the maps of the reconstructed convergence field. For comparison of the performance of the real space estimator, we also implement the harmonic space estimator by applying the procedure described in Appendix~\ref{app:harmonic_estimator_implementation}.
In this study we will be using two experiments inspired by the specifications of the PLANCK experiment for the $\nu=143~{\rm GHz}$ channel \cite{planck} and only differ in the detector noise. The experiments, denoted by {\designer} and {\planck,} both have $\theta_{fwhm}=7.8~{\rm arcmin}$ for the beam full-width at half maximum, and 
$\sigma_{pix}=0$ and $\sigma_{pix}=6.8~f_{sky}^{1/2}~\mu K/{\rm rad}$ for the white noise amplitude per beam width respectively, 
where $f_{sky}$ is the fraction area of the sky covered by the patch \cite{knox,tegmark}. 
We chose the experiments to have the same beam size in order not to entangle the effect of the noise with that of the beam. In future work we will study experiments with higher angular resolution such as ACT \cite{act} and SPT \cite{spt}.



\subsection{Generate the kernel}
The independence of the kernel eigenfunctions on the angular coordinate in the coordinates $(\boldsymbol{\ell}_{+},\boldsymbol{\ell}_{-})$ renders this parametrization more attractive to implement.
We start by defining the function $W(\boldsymbol{\ell}_{+},\boldsymbol{\ell}_{-})$ given by Eqn.~(\ref{eqn:w-q}). 
We then compute $W_m(\boldsymbol{\ell}_{+},\boldsymbol{\ell}_{-})$ using Eqn.~(\ref{eqn:w_m_ell}). The corresponding inverse Fourier transform $W_m(\boldsymbol{\theta}_{+},\boldsymbol{\theta}_{-})$ is computed using Eqn.~(\ref{eqn:w_m_theta}), with the final expression for the kernel following from Eqn.~(\ref{eqn:w_theta2}).

\begin{figure}[t]
\setlength{\unitlength}{1cm}
\vskip-2.5cm
\centerline{
\includegraphics[width=20cm]
{
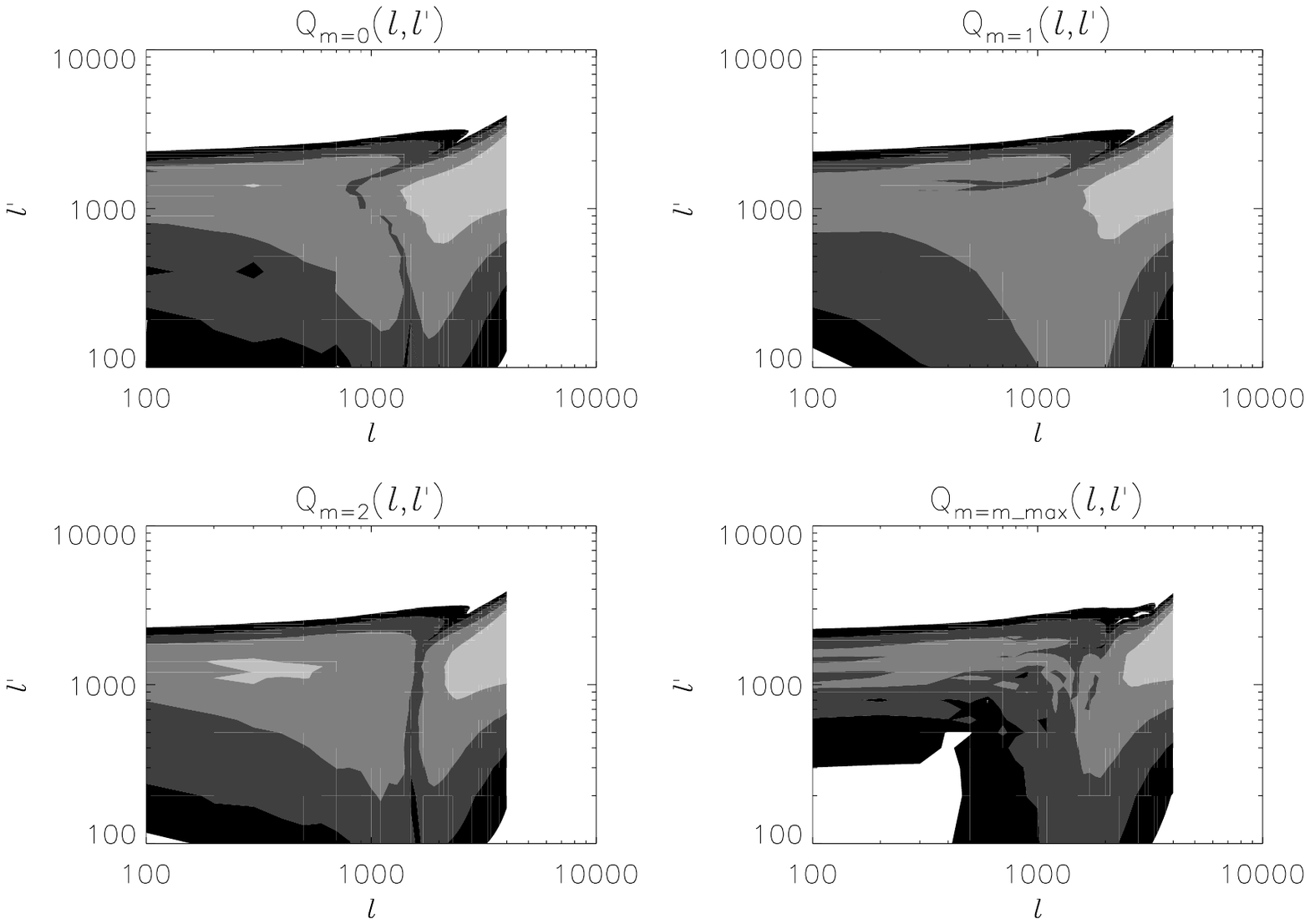}
}
\vskip-12.cm
\caption{\baselineskip=0.5cm{ 
{\bf Contour plots of $Q_{m}(\ell,\ell^{\prime}).$} 
The levels denote the orders of magnitude as fractions of the maximum value, delimited by 
$\{1,10^{-1},10^{-2},10^{-3},10^{-4}\}$ from the lightest to the darkest shade.
Here $m_{max}=16,$ $\ell_{max}=4000$ 
for PLANCK noise at $\nu=143~{\rm GHz}.$
}}
\label{fig:Q_m_ell_planck}
\end{figure}


\begin{figure}[t]
\setlength{\unitlength}{1cm}
\vskip-2.5cm
\centerline{
\includegraphics[width=20cm]
{
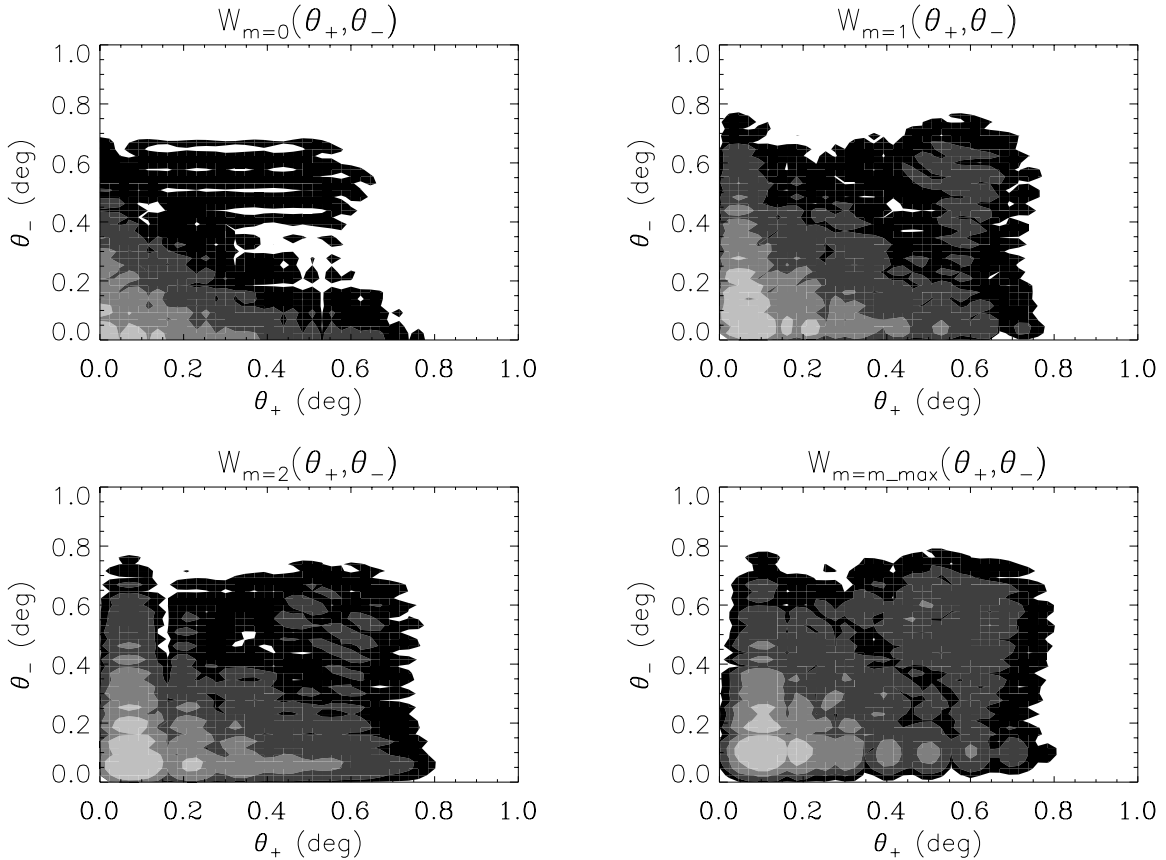}
}
\vskip-12.cm
\caption{\baselineskip=0.5cm{ 
{\bf Contour plots of $W_{m}(\theta_{+},\theta_{-}).$} 
The levels denote the orders of magnitude as fractions of the maximum value, delimited by 
$\{1,10^{-1},10^{-2},10^{-3},10^{-4}\}$ from the lightest to the darkest shade.
Here $m_{max}=4,$ $\ell_{max}=4000$  
for no detector noise.
}}
\label{fig:W_m_theta_designer}
\end{figure}

\begin{figure}[t]
\setlength{\unitlength}{1cm}
\vskip-2.5cm
\centerline{
\includegraphics[width=20cm]
{
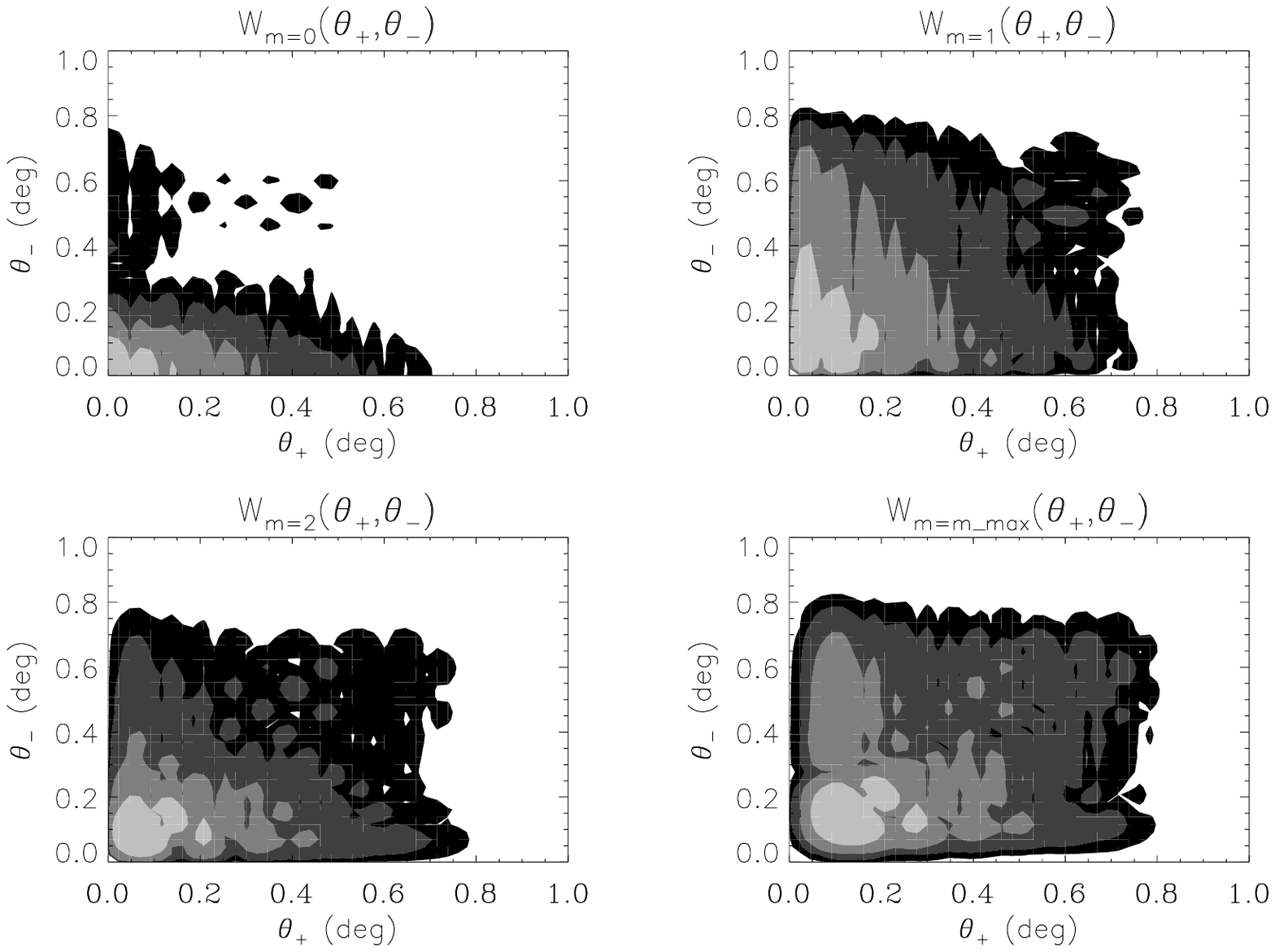}
}
\vskip-12.cm
\caption{\baselineskip=0.5cm{ 
{\bf Contour plots of $W_{m}(\theta_{+},\theta_{-}).$} 
The levels denote the orders of magnitude as fractions of the maximum value, delimited by 
$\{1,10^{-1},10^{-2},10^{-3},10^{-4}\}$ from the lightest to the darkest shade.
Here $m_{max}=4,$ $\ell_{max}=4000$ 
for PLANCK noise at $\nu=143~{\rm GHz}.$
}}
\label{fig:W_m_theta_planck}
\end{figure}

Two parameters that we need to define for the computation of the kernel that affect the reconstruction of the lensing convergence are the kernel extent in $\ell$ space, $\ell_{max},$ and the number of eigenfunctions at which we will truncate the series in $m,$ $m_{max}.$
Since the kernel is computed from the lensed map, 
the value of $\ell_{max}$ should be determined by the range in $\ell$ space that the experiment can probe before the beam and noise cut off the signal. We thus set $\ell_{max}$ equal to the higher $\ell$ moment that we use to synthesise the maps, i.e. $\ell_{max}=4000.$
This is also consistent with the value read off of $Q_{m}(\ell,\ell^{\prime})$ which we plot in Fig.~\ref{fig:Q_m_ell_planck}. \footnote{The small scale structure seen in Fig.~\ref{fig:Q_m_ell_planck} is likely numerical noise due to the discretization. Various tests that we tried seem to indicate that the quality of the lensing reconstruction is not significantly affected by this small scale structure. Nevertheless, we plan to improve the numerical calculation of the kernel and its transformation in future higher precision studies.}
We have tried using higher values for $\ell_{max}$ and found that neither the kernel nor the reconstructed power spectra changed significantly due to the fact that most of the lensing information is accessed around the beam scale. 
In the absence of detector noise, the kernel is displaced to higher values of $\ell$ and $\ell^{\prime},$ up to the chosen $\ell_{max}.$
To fix the value of $m_{max}$ we compared the relative amplitude of each eigenfunction $W_{m}(\boldsymbol{\theta}_{+},\boldsymbol{\theta}_{-})$
and discarded those that were less than $1\%$ of the dominant $m=0$ eigenfunction.
This corresponds to $m_{max}=4.$ 

For each experiment we generate $W_{m}(\theta_{+},\theta_{-})$ for the selected values of these two parameters.
In Figs.~\ref{fig:W_m_theta_designer} and \ref{fig:W_m_theta_planck}
we show the contour plots of four eigenfunctions of the kernel in real space for the {\designer} and the {\planck} experiment respectively. From the plot of the $m=0$ eigenfunction (which we have found to dominate the lensing reconstruction) we read off the value for the spatial extent of the kernel $\theta_{max},$ used in the convolution routine to compute the convergence map. In harmonic space the kernel peaks at the scale of the beam but has support at smaller $\ell$ values. The spatial extent of the kernel is thus larger than the beam scale and scales roughly linearly with the beam size.
We observe that the kernel is compact and extends only up to $0.70^{\circ}$ and $0.60^{\circ}$ at the $1\%$ level in the absence and in the presence of detector noise respectively. 
In practice, the cut-off in the kernel was tested by comparing the power spectrum of the convergence map reconstructed for different values of $\theta_{max}.$ For each experiment, we verified that increasing the value of $\theta_{max}$ beyond $0.35^{\circ}$ did not improve the reconstruction of the power spectrum, whereas reducing the value of $\theta_{max}$ below $0.35^{\circ}$ degraded the reconstruction. We thus set $\theta_{max}=0.35^{\circ}.$ 


\subsection{Compute the convergence map}
\begin{figure}[t]
\setlength{\unitlength}{1.cm}
\includegraphics[width=10cm]{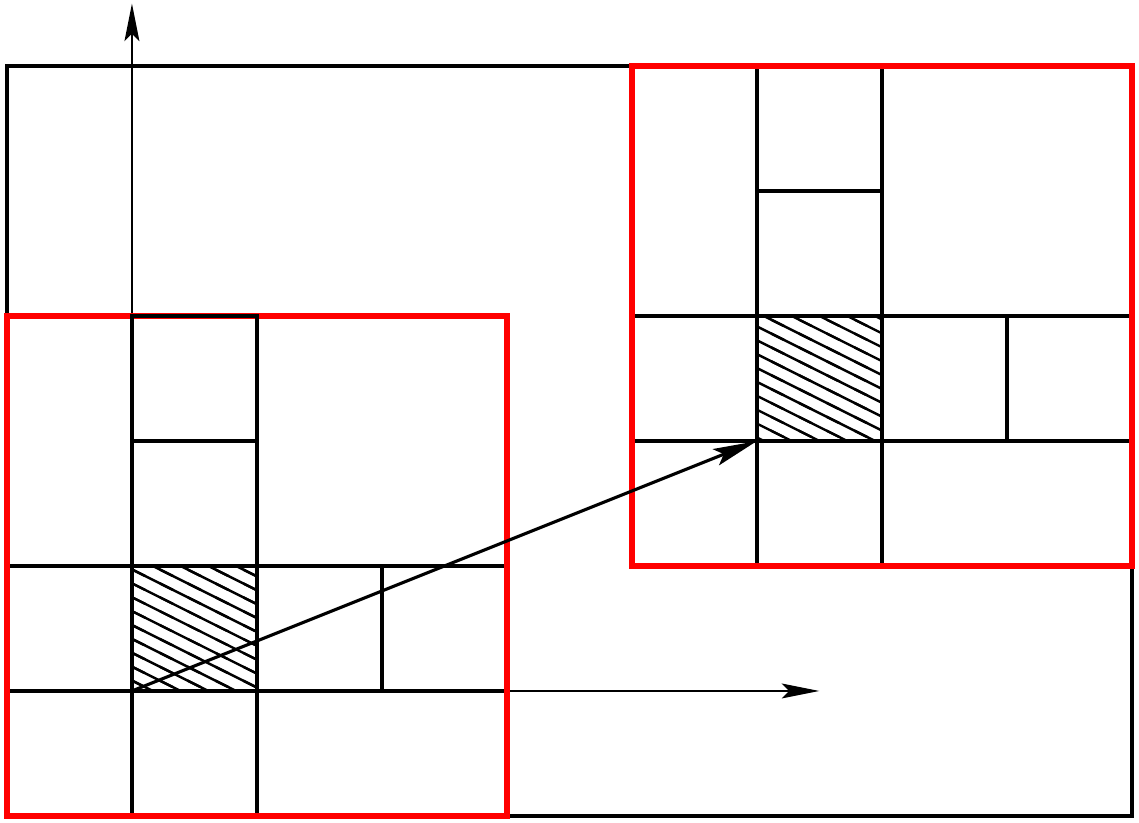}
\put(-9.6,6.8){$\boldsymbol{\theta}_{-}$}
\put(-5.2,2){$\boldsymbol{\theta}_{0}$}
\put(-3.3,0.6){$\boldsymbol{\theta}_{+}$}
\caption{\baselineskip=0.5cm{
{\bf Relative sizes in the reconstruction of the convergence map.}
The outer black square represents the map and the inner, dashed square centred at the origin represents the kernel. The red square around the kernel encompasses the points on the map which  contribute to the convergence on the point where the kernel is centred. The kernel moves over the map by successive translations indicated by the vector $\boldsymbol{\theta}_{0}.$ At the new centre, a new red square is defined. This procedure is repeated until the red square reaches the boundary of the map.
}}
\label{fig:pikassoyla_kernel}
\end{figure}

We compute the estimator by convolving the kernel with the lensed CMB map according to Eqn.~(\ref{eqn:kappa_w2}).
We implement the convolution using the discrete version of this equation as follows
\ba
\kappa_{0}(\boldsymbol{\theta}=0)
&=&
\sum_{m=-m_{max}}^{m_{max}}~
\sum _{\theta_{+x}}\sum _{\theta_{+y}}
 \sum _{\theta_{-x}}\sum _{\theta_{-y}}~
  \exp\left[im\arctan\left(
  {{\theta_{+y}-\theta_{-y}}\over {\theta_{+x}-\theta_{-x}}} \right) \right]\cr
&&\times
  \tilde T(\theta_{+x}+\theta_{-x},\theta_{+y}+\theta_{-y})~
  \tilde T(\theta_{+x}-\theta_{-x},\theta_{+y}-\theta_{-y})\cr
&&\times 
  W_{m}\left(\sqrt{\theta_{+x}^2+\theta_{+y}^2}, 
  \sqrt{\theta_{-x}^2+\theta_{-y}^2}\right).
\ea
The size of the kernel determines the extent of the temperature map over which the convolution is computed.
Since the temperature map is evaluated at the sum and difference of the contributing pixels, we need a region of the temperature map that is four times the length of the kernel to compute the convolution (see Fig.~\ref{fig:pikassoyla_kernel}).
At an arbitrary point $\boldsymbol{\theta}=(\theta_{0x},\theta_{0y})$ the convergence is given by the estimator at the origin translated by the vector $\boldsymbol{\theta}.$

We then average the reconstructed convergence map over all pixels to extract the zero mode, which is then subtracted from the map as discussed in the text following Eqn.~(\ref{eqn:kappa_w3}).
Finally we subtract the convergence power spectrum of a Monte Carlo realization of one hundred unlensed CMB maps from the convergence power spectrum of the lensing potential, thus removing the gaussian bias from the estimated power spectrum. 
Our implementation of the real space estimator takes a little over one minute in total on a desktop computer, including the calculation of the  kernel (about 30~sec) and the reconstruction of the convergence map (about 40~sec). This is about six times the time that the harmonic space implementation takes to reconstruct the lensing field.

\section{Results and Discussion}
\label{sec:discussion}

\begin{figure}[t]
\setlength{\unitlength}{1cm}
\vskip-1.5cm
\centerline{
\hskip-0.5cm
\includegraphics[width=12cm]{
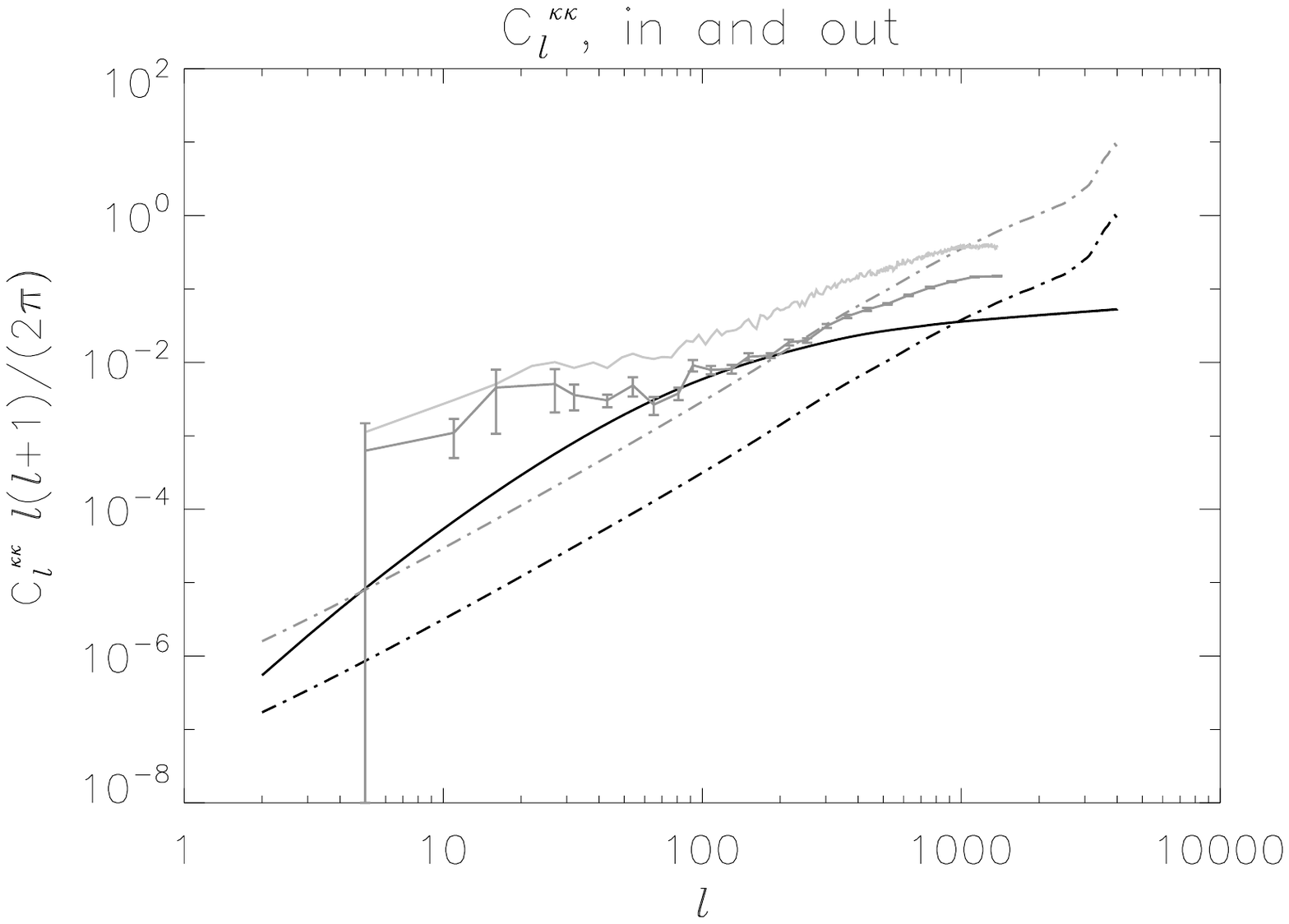}
\hskip-3cm
\includegraphics[width=12cm]{
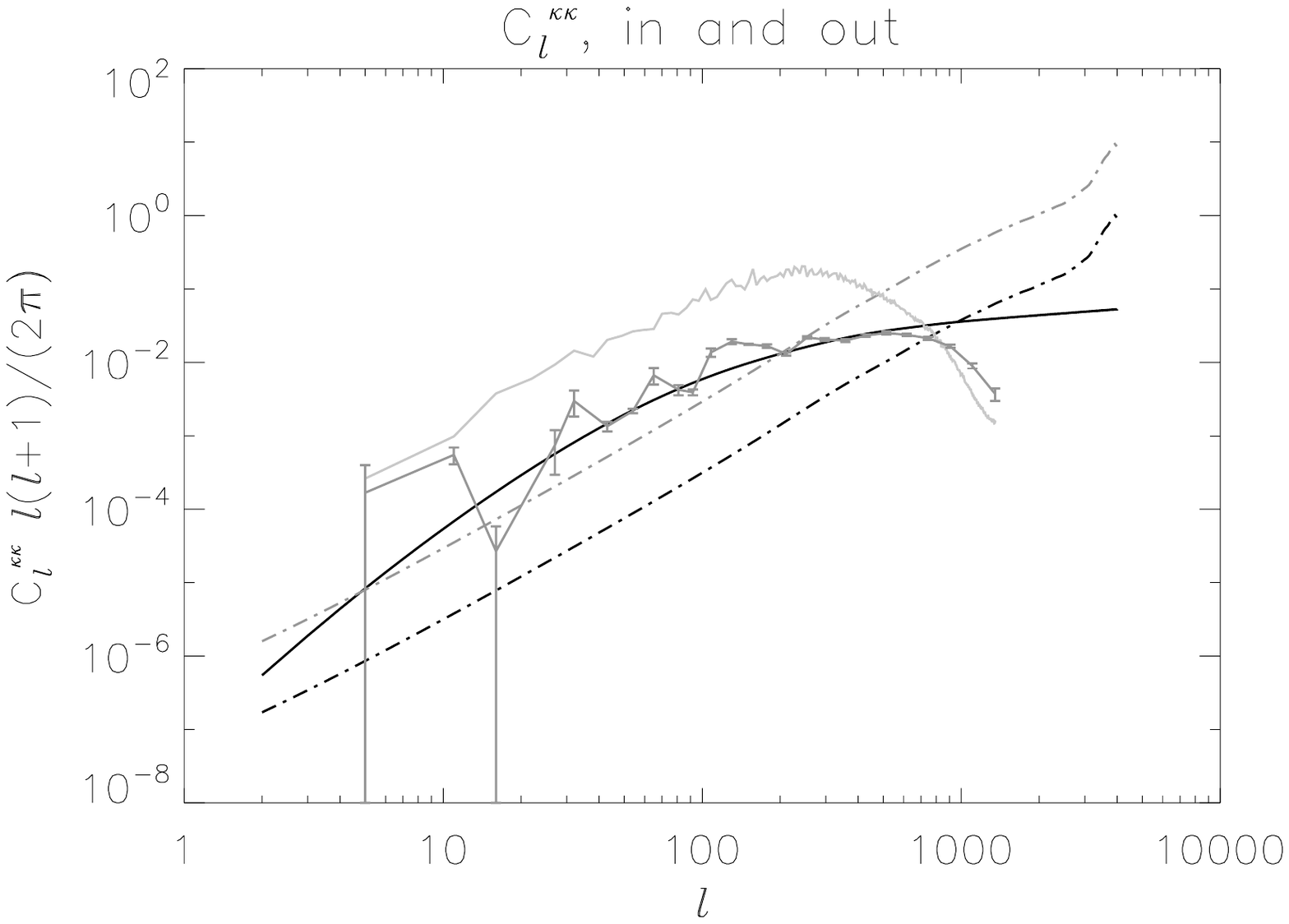}
} 
\vskip-7.5cm
\caption{\baselineskip=0.5cm{ 
{\bf Mean square of the convergence map reconstructed with the harmonic space estimator (left panel) and the real space estimator (right panel).} 
The smooth black curve is the input power spectrum. The light grey and dark grey solid lines are the output power spectra before and after the removal of the bias respectively. 
The error bars measure the total standard deviation binned over logarithmically spaced intervals in $\ell.$ 
Here $\theta_{fwhm}=7.8',$ $FOV_{map}=66.6^{\circ}$ and 
$m_{max}=4,$ $\ell_{max}=4000$ for no detector noise.
}}
\label{fig:cl_kk2_designer}
\end{figure}

\begin{figure}[h]
\setlength{\unitlength}{1cm}
\vskip-1.5cm
\centerline{
\hskip-0.5cm
\includegraphics[width=12cm]
{
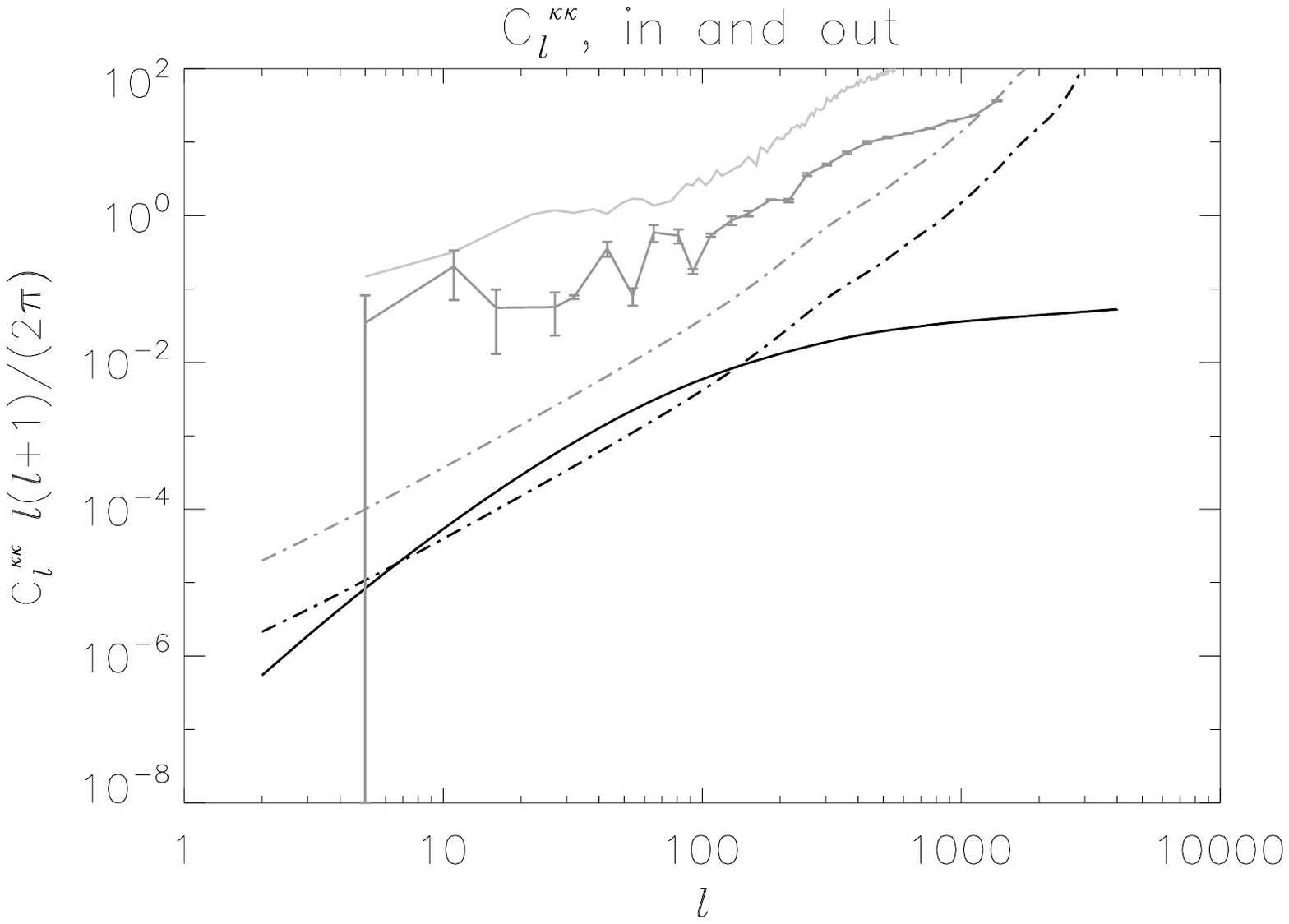}
\hskip-3cm
\includegraphics[width=12cm]
{
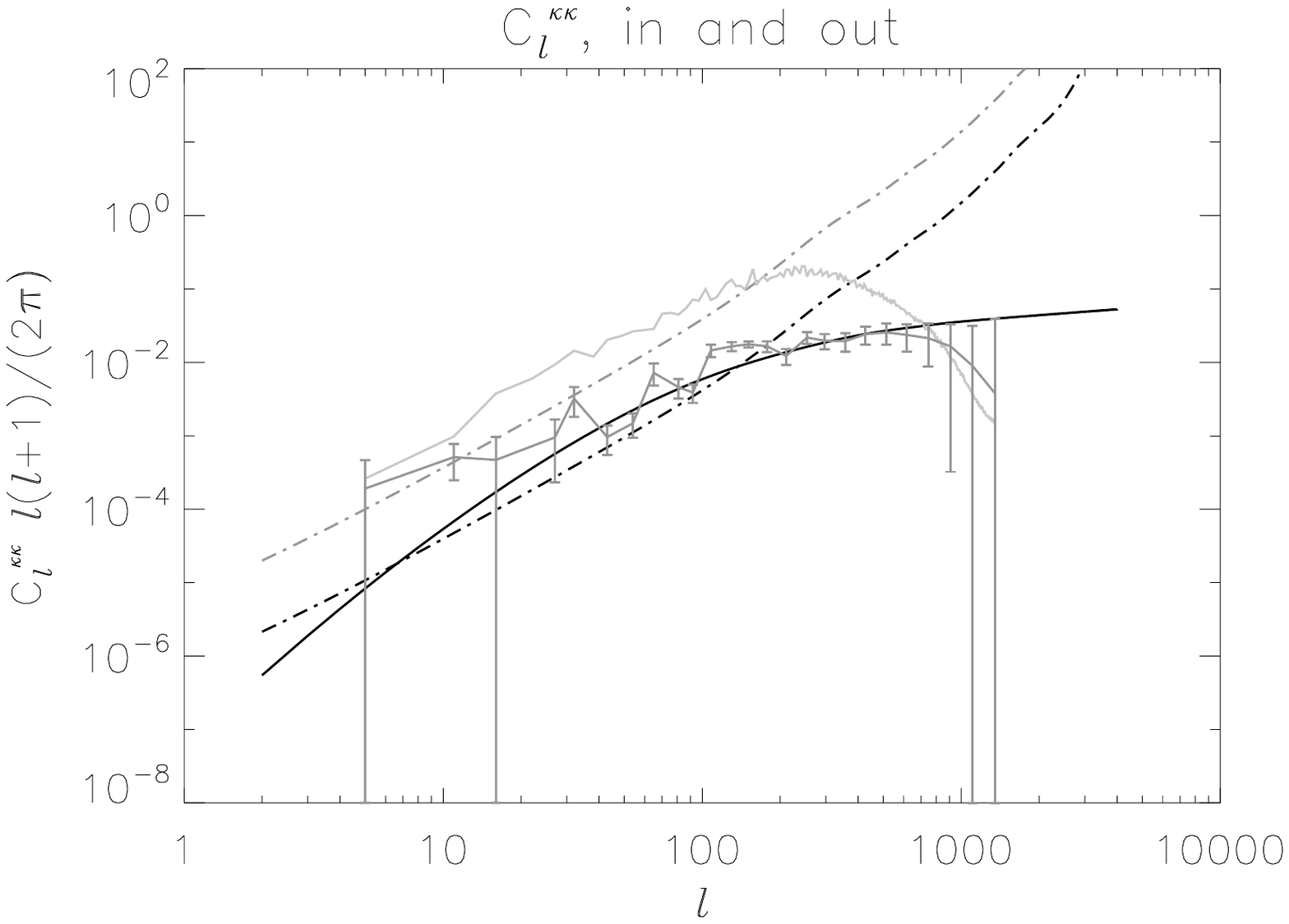}
}
\vskip-7.5cm
\caption{\baselineskip=0.5cm{ 
{\bf Mean square of the convergence map reconstructed with the harmonic space estimator (left panel) and the real space estimator (right panel).} 
The smooth black curve is the input power spectrum. The light grey and dark grey solid lines are the output power spectra before and after the removal of the bias respectively. 
The error bars measure the total standard deviation binned over logarithmically spaced intervals in $\ell.$ 
Here $\theta_{fwhm}=7.8',$ $FOV_{map}=66.6^{\circ},$ and 
$m_{max}=4,$ $\ell_{max}=4000$
for PLANCK noise at $\nu=143~{\rm GHz}.$
}}
\label{fig:cl_kk2_planck}
\end{figure}

For each experiment 
we plot the output power spectrum $C_{\ell}^{\kappa_{out}\kappa_{out}}$, both before (light grey solid line) and after (dark grey solid line) the removal of the bias, against the input power spectrum $C_{\ell}^{\kappa_{in}\kappa_{in}}$ (black solid line).
In Figs.~\ref{fig:cl_kk2_designer} and \ref{fig:cl_kk2_planck}
we plot the reconstructed power spectra for the {\designer} and the {\planck} experiment respectively.
We also plot the variance of the estimator 
${\rm Var}[\hat\kappa_{0}(\boldsymbol{\ell})]$
(black dot-dashed line) 
as well as 
${\rm Var}[\hat\kappa_{0}(\boldsymbol{\ell})]/f_{sky}$ 
(grey dot-dashed line) as an auxiliary tool for visualization.  
The error bars on the recovered power spectra are computed from the 
standard deviation of the reconstructed power spectrum, binned over logarithmically spaced intervals in $\ell$ space and added in quadrature.
The variance of the reconstructed power spectrum encompasses both the variance of the optimal estimator and the average value of the convergence fluctuation in the map, weighted by the fraction of the area of the sky covered by the patch, $f_{sky},$ as follows \cite{tegmark}
\ba
{\rm Var}[C_{\ell}^{\kappa_{0}\kappa_{0}}]
\approx{1\over {\ell~f_{sky}}}\left( {\rm Var}[\hat \kappa_{0}(\boldsymbol{\ell})]+C_{l}^{\kappa_{0}\kappa_{0}}\right)^2.
\ea
We observe that the degree of recovery of the input power spectra is not the same throughout the interval in $\ell.$ In particular, 
at small $\ell$ the cosmic variance dominates the error of the reconstruction, whereas at large $\ell$ the variance of the estimator takes over.
The region where 
$C_{\ell}^{\kappa_{in}\kappa_{in}}>
{\rm Var}[\hat\kappa_{0}(\boldsymbol{\ell})]$
indicates the interval in $\ell$ where the lensing potential can in theory be mapped. 
Accordingly, this is also the region where the reconstructed convergence power spectra tracks more closely the input power spectrum. Also visible are the fluctuations in the recovered power spectrum which arise from the subtraction of the noise bias term estimated from a finite number of realizations of the unlensed CMB map.
For the {\planck} experiment, using the real space implementation we achieve a reasonable reconstruction of the input power spectrum in the interval where the input signal is larger than the variance of the estimator, although on smaller scales, $\ell > 600,$ there is a decrease in the recovered power. Comparing the reconstructed power spectra for the {\designer} and {\planck} experiments we observe that the real space implementation is fairly insensitive to the detector noise and returns a reconstruction consistent with the input power spectrum. We can understand these results as follows.

As attested by the power spectra of the maps reconstructed with the real space estimator, there is a loss of power at small scales beginning around the angular scale corresponding to the size of the kernel.
This loss of power is a consequence of averaging modes smaller than the finite extent of the kernel
and hence is an intrinsic constraint of the real space implementation.
This intrinsic constraint, 
studied in more detail in Ref.~\cite{bucher2010}, can be overcome by shrinking the extent of the kernel in real space, which can be achieved with a smaller beam and detector noise, thereby moving the support for the kernel to larger $\ell.$
The $\ell$ interval over which the power spectrum can be recovered by the real space estimator is determined by the characteristic lengths of the map and the kernel as follows. The lower $\ell$ limit is determined by the size of the map, since it measures the longest wavelength mode that can be enclosed within the size of the map. The upper $\ell$ limit is determined by the size of the kernel, since it measures the smallest wavelength mode that the kernel can probe, below which there is a loss of power due to the averaging of modes smaller than the size of the kernel.
Within this $\ell$ range, the real space estimator seems to perform fairly well both in the absence and in the presence of detector noise. Note that the $\ell$ range for a reasonable  reconstruction with the harmonic space estimator is wider, being limited at higher $\ell$ by the angular scale corresponding to the beam size, since this is the scale which constrains the action of the kernel in harmonic space.

\begin{figure}[t]
\setlength{\unitlength}{1cm}
\vskip-1.5cm
\centerline{
\hskip-0.5cm
\includegraphics[width=12cm]
{
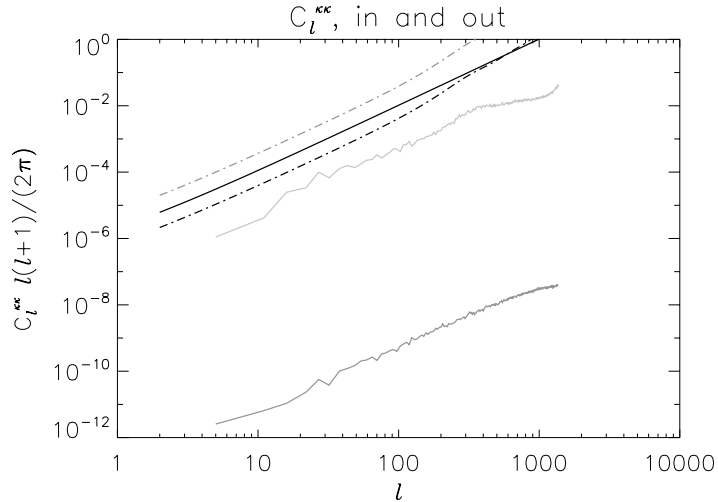}
}
\vskip-7.5cm
\caption{\baselineskip=0.5cm{ 
{\bf Mean square of the convergence map reconstructed with the harmonic space estimator (light grey line) and the real space estimator (dark grey line) from a pure white noise map.} 
The grey solid lines are the output power spectra whereas the smooth black curve is the input power spectrum.
Here $\theta_{fwhm}=7.8',$ $FOV_{map}=66.6^{\circ},$ and 
$m_{max}=4,$ $\ell_{max}=4000$
for PLANCK noise at $\nu=143~{\rm GHz}.$
}}
\label{fig:cl_kk2_planck_pure_noise}
\end{figure}

Comparing the reconstructed power spectra in the absence and in the presence of detector noise, we observe that the real space estimator appears to be insensitive to the experimental noise. 
This is because the estimate of the convergence in each pixel is given by the sum of the product of pairs of neighbouring pixels weighted by the kernel. 
As a result, the noise, being independent in each pixel, is averaged out. 
To substantiate this result, we reconstructed the convergence map from a pure white noise input map using the real space estimator and the harmonic space estimator. In Fig.~\ref{fig:cl_kk2_planck_pure_noise} we observe that the power spectrum recovered by the real space estimator is about seven orders of magnitude smaller than the input power spectrum, whereas 
the power spectrum recovered by the harmonic space estimator is comparable to the input power spectrum. 

Despite our attempt to remove the CMB and the detector noise bias from the reconstructed convergence power spectrum, there still remains a bias in the harmonic space implementation, as observed in Figs.~\ref{fig:cl_kk2_designer} and \ref{fig:cl_kk2_planck}. This bias presumably arises from a coupling between unlensed temperature modes due to the finite size of the map, providing an additional source of lensing which we have not treated here.
This mode coupling is absent in the reconstruction over the full sky \cite{inpaint}. The real space implementation appears to be insensitive to this bias.




\section{Summary and future work}
\label{sec:summary}

In this paper we 
implemented a new estimator of the convergence field 
for the extraction of the lensing potential from CMB maps.
Our interest was in reconstructing the convergence field from CMB maps from which contaminants, such as point sources and the SZ effect, 
had previously been removed using multi-frequency information. 

The new estimator 
acts locally in real space, thus 
being able to treat the excision of pixels and nonuniform sky coverage in a flexible manner.
We implemented 
the estimator on two experimental setups, one without and one with detector noise, based on the specifications of  PLANCK for the $\nu=143~{\rm GHz}$ channel. 
For comparison of the performance of the new estimator, we also implemented the conventional estimator which acts in harmonic space.
From the two experiments we learned that by increasing the range of the kernel in $\ell$ space, and consequently reducing its range in $\theta$ space, we improve the reconstruction of the power spectrum at small scales.
The finite extent of the kernel is a desirable property of the proposed estimator, since in theory it allows the reconstruction of the lensing convergence, manifested essentially at very small scales, from a small map of the sky as long as the kernel probes sufficiently small angular scales. 
Even though there is a loss of power on scales smaller than the finite extent of the kernel, this effect could be studied using simulations to determine a form factor that could be applied to correct the loss of power \cite{bucher2010}.
We also observed that although the real space estimator is limited, in a statistical sense, to be as good as the harmonic space estimator, 
it has the advantage over the harmonic space estimator of being
less sensitive to white noise. In practice, however, CMB experiments have to deal with correlated noise, which will pose as much of a challenge to the real space estimator as to the harmonic space estimator, most likely requiring the construction of a more carefully designed kernel in harmonic space that downweights the correlated modes before transforming to real space.


On the basis of the results presented here, we envisage two follow-up studies.
In the first study we intend to 
optimize the current implementation of the real space estimator to use a kernel capable of probing smaller angular scales 
in the reconstruction of the lensing convergence, as we would expect for the ACT and SPT experiments.
As we have demonstrated, the real space estimator can be applied to small patches of sky without incurring in the serious spectral leakage that affects the harmonic space estimator on rectangular, rather than torodial, domains in the presence of red spectra. In real space, the local filter acts up to a kernel length away from the edge of the map without leakage of power. We also intend to study the effect of a mask due to the excision of bad pixels, e.g. arising from the removal of bright point sources, that can cause aliasing of power from large scales to small scales. Inpainting  has been proposed as a means of interpolating across masked regions that preserves the statistical properties of the map \cite{inpaint}, so we plan to investigate how sensitive the real space estimator is to complex masks with and without the use of such techniques.

A straightforward extension will be to develop the analogous estimator for the shear components of the convergence tensor and work out how to combine the different components for the reconstruction of the lensing potential. Since the shear components do not contain the $\ell=0$ mode, the corresponding estimators will be less sensitive to a bias. The dilatation and shear components provide complementary information on the lensing potential but are not independent, being related by consistency conditions that are algebraic in harmonic space. Since these components probe the monopole and quadrupole sectors respectively, the noise associated with these components will be uncorrelated and the reconstruction can be harmonized, with the noise being reduced via inverse variance weighting of these components. This point is discussed in more detail in Ref.~\cite{bucher2010}.

In the second study we intend to extend the implementation of the real space estimator to CMB polarisation so as to optimize the reconstruction of the lensing potential from the PLANCK data.

\vspace{0.6cm}
\centerline{\bf {Acknowledgments}}
The authors are supported by the South African National Research Foundation.
The authors acknowledge the use of codes by M~Bucher and the use of CAMB \cite{camb}.
The authors also thank M~Bucher, S~Das, D~Spergel, B Sherwin and M~Remazeilles for useful discussions, as well as the anonymous referee for a number of useful comments. CSC also acknowledges the hospitality of the Astrophysical Sciences Department, Princeton University.
\hfill

\appendix


\section{Derivation of the harmonic space estimator}
\label{app:harmonic_estimator_description}

In this Appendix we derive the optimal estimator for the lensing potential in harmonic space.
From Eqn.~(\ref{eqn:c_tt}) a crude estimator for the lensing potential would be 
the $\boldsymbol{\ell}\not =0$ term
\ba
\hat \psi(\boldsymbol{\ell},\boldsymbol{\ell}^{\prime})
={1\over (2\pi)^2}{
\tilde T(\boldsymbol{\ell}^{\prime})~
\tilde T(\boldsymbol{\ell}-\boldsymbol{\ell}^{\prime})
\over \left[ \boldsymbol{\ell}\cdot\boldsymbol{\ell}^{\prime}~
C_{\ell^{\prime}}
+\boldsymbol{\ell}\cdot(\boldsymbol{\ell}-\boldsymbol{\ell}^{\prime})~
C_{\vert \boldsymbol{\ell}-\boldsymbol{\ell}^{\prime}\vert}\right]}.
\ea
This way we would be estimating the anisotropic component of the convergence which is due to the lensing potential only, i.e. without the isotropic zero mode.

We now seek for the optimal combination 
\ba
\hat \psi(\boldsymbol{\ell})
=\int {d^2\boldsymbol{\ell}^{\prime}\over (2\pi)^2}~
\hat \psi(\boldsymbol{\ell},\boldsymbol{\ell}^{\prime})~
G(\boldsymbol{\ell},\boldsymbol{\ell}^{\prime}),
\ea
where the optimal weights $G(\boldsymbol{\ell},\boldsymbol{\ell}^{\prime})$ 
are given by
\ba
G(\boldsymbol{\ell},\boldsymbol{\ell}^{\prime})
={1\over \sigma_{\hat \psi(\boldsymbol{\ell},\boldsymbol{\ell}^{\prime})}^{2}}
\left[
\int {d^2\boldsymbol{\ell}\over (2\pi)^2} 
\int {d^2\boldsymbol{\ell}^{\prime}\over (2\pi)^2}
{1\over \sigma_{\hat \psi(\boldsymbol{\ell},\boldsymbol{\ell}^{\prime})}^{2}}\right]^{-1}
\ea
so as to minimize the variance,
and are such that
\ba
\int {d^2\boldsymbol{\ell}\over (2\pi)^2} 
\int {d^2\boldsymbol{\ell}^{\prime}\over (2\pi)^2}~
G(\boldsymbol{\ell},\boldsymbol{\ell}^{\prime})=1
\ea
so as to make the estimator unbiased 
$\bigl<\hat \psi(\boldsymbol{\ell},\boldsymbol{\ell}^{\prime})\bigr>=\psi(\boldsymbol{\ell}).$
Here $\sigma_{\hat \psi(\boldsymbol{\ell},\boldsymbol{\ell}^{\prime})}^{2}$ is the variance of  $\hat \psi(\boldsymbol{\ell},\boldsymbol{\ell}^{\prime})$
which for the case that $\psi(\boldsymbol{\ell})$ is centred at zero becomes
\ba
\sigma_{\hat \psi(\boldsymbol{\ell},\boldsymbol{\ell}^{\prime})}^{2}
&=&\left<
\left( 
\hat \psi(\boldsymbol{\ell}_{1},\boldsymbol{\ell}^{\prime}_{1})
-\psi(\boldsymbol{\ell}_{1})\right)
\left(
\hat \psi(\boldsymbol{\ell}_{2},\boldsymbol{\ell}^{\prime}_{2})
-\psi(\boldsymbol{\ell}_{2})\right)^{\ast}
\right> \cr
&=&{1\over (2\pi)^4}
\left<
\tilde T(\boldsymbol{\ell}^{\prime}_{1})~
\tilde T(\boldsymbol{\ell}_{1}-\boldsymbol{\ell}^{\prime}_{1})~
\tilde T^{\ast}(\boldsymbol{\ell}^{\prime}_{2})~
\tilde T^{\ast}(\boldsymbol{\ell}_{2}-\boldsymbol{\ell}^{\prime}_{2})
\right>\cr
&&\times
{1
\over \left[ \boldsymbol{\ell}_{1}\cdot\boldsymbol{\ell}^{\prime}_{1}~
C_{\ell^{\prime}_{1}}
+\boldsymbol{\ell}_{1}\cdot(\boldsymbol{\ell}_{1}-\boldsymbol{\ell}^{\prime}_{1})~
C_{\vert \boldsymbol{\ell}_{1}-\boldsymbol{\ell}^{\prime}_{1}\vert}\right]}
{1
\over \left[ \boldsymbol{\ell}_{2}\cdot\boldsymbol{\ell}^{\prime}_{2}~
C_{\ell^{\prime}_{2}}
+\boldsymbol{\ell}_{2}\cdot(\boldsymbol{\ell}_{2}-\boldsymbol{\ell}^{\prime}_{2})~
C_{\vert \boldsymbol{\ell}_{2}-\boldsymbol{\ell}^{\prime}_{2}\vert}\right]}
\cr
&=&2~\delta(\boldsymbol{\ell}_{1}-\boldsymbol{\ell}_{2})~\delta(\boldsymbol{\ell}^{\prime}_{1}-\boldsymbol{\ell}^{\prime}_{2})
{{
\tilde C_{\ell^{\prime}_{1}}
\tilde C_{\vert \boldsymbol{\ell}_{1}-\boldsymbol{\ell}^{\prime}_{1}\vert} } 
\over{
\left[ \boldsymbol{\ell}_{1}\cdot\boldsymbol{\ell}^{\prime}_{1}~
C_{\ell^{\prime}_{1}}
+\boldsymbol{\ell}_{1}\cdot(\boldsymbol{\ell}_{1}-\boldsymbol{\ell}^{\prime}_{1})~
C_{\vert \boldsymbol{\ell}_{1}-\boldsymbol{\ell}^{\prime}_{1}\vert}\right]^2
}}.
\ea
We then find for the optimal estimator that
\ba
\hat \psi(\boldsymbol{\ell})
=\mathcal{N_{\ell}}
\int {d^2\boldsymbol{\ell}^{\prime}\over (2\pi)^2}~
\tilde T(\boldsymbol{\ell}^{\prime})~
\tilde T(\boldsymbol{\ell}-\boldsymbol{\ell}^{\prime})~
{1\over 2}{
{\boldsymbol{\ell}\cdot\boldsymbol{\ell}^{\prime}~
C_{\ell^{\prime}}
+\boldsymbol{\ell}\cdot(\boldsymbol{\ell}-\boldsymbol{\ell}^{\prime})~
C_{\vert \boldsymbol{\ell}-\boldsymbol{\ell}^{\prime}\vert}}
\over
\tilde C_{\ell^{\prime}}
\tilde C_{\vert \boldsymbol{\ell}-\boldsymbol{\ell}^{\prime}\vert} },
\ea
where 
\ba
{1\over \mathcal{N}_{\ell}}
=\int {d^2\boldsymbol{\ell}^{\prime}\over (2\pi)^2}~
{1\over 2}
{ 
{[ \boldsymbol{\ell}\cdot\boldsymbol{\ell}^{\prime}~
C_{\ell^{\prime}}
+\boldsymbol{\ell}\cdot(\boldsymbol{\ell}-\boldsymbol{\ell}^{\prime})~
C_{\vert \boldsymbol{\ell}-\boldsymbol{\ell}^{\prime}\vert} ]^2}
 \over {
\tilde C_{\ell^{\prime}}~
\tilde C_{\vert \boldsymbol{\ell}-\boldsymbol{\ell}^{\prime}\vert}}
}.
\ea
Note that the calculation of the variance presupposes a prior knowledge of the unlensed power spectrum of the temperature anisotropies. Here we use the same $TT$ power spectrum that we use for the synthesis of the CMB map.
 
The error associated with the reconstruction of the lensing potential with this estimator is given by the variance of the estimator. 
We can identify the variance of the estimator with the power spectrum of the estimator
as follows $\sigma_{\hat \psi(\boldsymbol{\ell},\boldsymbol{\ell}^{\prime})}^{2}
=\bigl<\hat \psi(\boldsymbol{\ell}_{1},\boldsymbol{\ell}^{\prime}_{1})~
\hat \psi^{\ast}(\boldsymbol{\ell}_{2},\boldsymbol{\ell}^{\prime}_{2})\bigr>=(2\pi)^2~\delta(\boldsymbol{\ell}_{1}-\boldsymbol{\ell}_{2})~
\delta(\boldsymbol{\ell}^{\prime}_{1}-\boldsymbol{\ell}^{\prime}_{2})~
{\rm Var}[\hat \psi(\boldsymbol{\ell},\boldsymbol{\ell}^{\prime})].$ 
It follows that the variance of the optimal estimator is given by
\ba
\sigma^{2}_{\hat\psi(\boldsymbol{\ell})}=
\left[ 
\int {d^2\boldsymbol{\ell}^{\prime}_{2}\over (2\pi)^2}
\int {d^2\boldsymbol{\ell}^{\prime}_{1}\over (2\pi)^2}
{1\over \sigma_{\hat \psi(\boldsymbol{\ell},\boldsymbol{\ell}^{\prime})}^{2}}\right]^{-1}
= (2\pi)^2~\delta(\boldsymbol{\ell}_{1}-\boldsymbol{\ell}_{2})~
{\rm Var}[\hat \psi(\boldsymbol{\ell})]
\ea
where
\ba
{\rm Var}[\hat \psi(\boldsymbol{\ell})]
=\left[ 
\int {d^2\boldsymbol{\ell}^{\prime}\over (2\pi)^2}~
{1\over 2}
{ 
{[ \boldsymbol{\ell}\cdot\boldsymbol{\ell}^{\prime}~
C_{\ell^{\prime}}
+\boldsymbol{\ell}\cdot(\boldsymbol{\ell}-\boldsymbol{\ell}^{\prime})~
C_{\vert \boldsymbol{\ell}-\boldsymbol{\ell}^{\prime}\vert} ]^2}
 \over {
\tilde C_{\ell^{\prime}}~
\tilde C_{\vert \boldsymbol{\ell}-\boldsymbol{\ell}^{\prime}\vert}}
}
\right]^{-1}
=\mathcal{N}_{\ell}.
\label{eqn:variance}
\ea
The variance gives an upper limit to the dispersion about the mean reconstructed map of the output maps reconstructed with the estimator from a finite sample of input synthesised maps,
thus being a measure that can be used to validate the estimator.

The expression for the optimal estimator can most straightforwardly be modified to  incorporate detector noise and finite beam width by replacing $\tilde C_{\ell^{\prime}}$
by $\tilde C_{\ell^{\prime}}+N_{\ell^{\prime}}.$
The noise power spectrum $N_{\ell}$ 
is given by the inverse-sum over the number of channels of the detector noise
$n_{i}(\ell)$ of each channel $i$ 
\ba
N_{\ell}=
\left( \sum_{i=0}^{\tt num\_chann} 
{1\over n_{i}(\ell)} \right)^{-1}.
\label{eqn:n_ell}
\ea
The detector noise is modelled by a gaussian signal on the beam size 
and includes both the white noise amplitude and the beam profile attenuation factor as follows
\ba
n_{i}(\ell)
=\left({\theta_{fwhm}}_{i}{\sigma_{pix}}_{i}\right)^2
\exp[({\theta_{fwhm}}_{i})^2\ell(\ell+1)/(8\ln2)].
\ea
Here ${\theta_{fwhm}}_{i}$ is the beam full-width at half-maximum of the beam 
and
${\sigma_{pix}}_{i}$ 
is the white noise amplitude per beam width.

We can regard $N_{\ell}$ as the total variance of a multiple sampling of size the number of channels, each sampling $i$ having variance 
$n_{i}(\ell).$
Finally, the estimator for the lensing potential becomes
\ba
\hat \psi(\boldsymbol{\ell})
=\mathcal{N}_{\ell}
\int {d^2\boldsymbol{\ell}^{\prime}\over (2\pi)^2}~
\tilde T(\boldsymbol{\ell}^{\prime})~
 \tilde T(\boldsymbol{\ell}-\boldsymbol{\ell}^{\prime})~
{1\over 2}{
{\boldsymbol{\ell}\cdot\boldsymbol{\ell}^{\prime}~
C_{\ell^{\prime}}
+\boldsymbol{\ell}\cdot(\boldsymbol{\ell}-\boldsymbol{\ell}^{\prime})~
C_{\vert \boldsymbol{\ell}-\boldsymbol{\ell}^{\prime}\vert}}
\over { [\tilde C_{\ell^{\prime}}+N_{\ell}]
[\tilde C_{\vert \boldsymbol{\ell}-\boldsymbol{\ell}^{\prime}\vert}
 +N_{\vert \boldsymbol{\ell}-\boldsymbol{\ell}^{\prime}\vert}]}
},
\ea
where 
\ba
{1\over \mathcal{N}_{\ell} }
=\int {d^2\boldsymbol{\ell}^{\prime}\over (2\pi)^2}~ 
{1\over 2}
{ 
{[ \boldsymbol{\ell}\cdot\boldsymbol{\ell}^{\prime}~
C_{\ell^{\prime}}
+\boldsymbol{\ell}\cdot(\boldsymbol{\ell}-\boldsymbol{\ell}^{\prime})~
C_{\vert \boldsymbol{\ell}-\boldsymbol{\ell}^{\prime}\vert} ]^2}
 \over { [\tilde C_{\ell^{\prime}}+N_{\ell^{\prime}}]
 [\tilde C_{\vert \boldsymbol{\ell}-\boldsymbol{\ell}^{\prime}\vert}
 +N_{\vert \boldsymbol{\ell}-\boldsymbol{\ell}^{\prime}\vert}]}
}.
\label{eqn:norm_factor}
\ea

The estimator for the deflection angle 
$\boldsymbol{\alpha}(\boldsymbol{\theta})=\nabla\psi(\boldsymbol{\theta})$
and the convergence 
$\kappa_{0}(\boldsymbol{\theta})=-\nabla^2\psi(\boldsymbol{\theta})/2$
will be given respectively by the gradient and the divergence in real space of 
$\hat \psi(\boldsymbol{\ell}).$ 
In $\boldsymbol{\ell}$ space this amounts simply to 
\ba
\hat {\boldsymbol \alpha}(\boldsymbol{\ell})
&=&i\boldsymbol{\ell}~\hat \psi(\boldsymbol{\ell})\\
\hat \kappa_{0}(\boldsymbol{\ell})
&=&
\ell ^2~\hat \psi(\boldsymbol{\ell})
\ea
with variances
${\rm Var}[\hat\alpha(\boldsymbol{\ell})]
=\ell^2~{\rm Var}[\hat\psi(\boldsymbol{\ell})]$ 
and 
${\rm Var}[\hat\kappa_{0}(\boldsymbol{\ell})]
=
\ell^4~{\rm Var}[\hat\psi(\boldsymbol{\ell})]$ respectively.


\section{Synthesis of the lensed CMB map}
\label{app:map_synthesis}

For the construction of the CMB map and the lensing map we used 
power spectra generated by CAMB \cite{camb}, respectively the TT and the $\psi\psi$ power spectra, on scales with $\ell\le 4000.$
Given the nature of the CMB lensing and the form of the data to be analysed, we will be working upon a square patch, thus bounded, of the unbounded celestial sphere. 
For a proper treatment of the boundaries of the patch, we must simulate the unboundedness of the sphere by implementing periodic boundary conditions.
In order to reproduce periodic boundary conditions,
we apodized the CMB map by smoothing to zero over a strip around the boundary with a given width fraction (in this case 0.10). 
The requirement of periodic boundary conditions makes the implementation of the quadratic estimator in harmonic space easier but is not necessary for the implementation of the real space estimator.

We then applied the lensing map to the CMB map to produce the lensed CMB map. This consisted of shifting the CMB map by the lensing potential assuming the Born approximation, as detailed in the Appendix of Ref.~\cite{lewis}. 
A fast way to construct lensed maps to good accuracy is to synthesise 
the CMB map 
with a resolution higher (at least twice larger) than the resolution desired for the lensed CMB map 
and then remap the pixels of the CMB map by the deflection angle onto the lens plane.
The size of a map is given by the pixel size times the number of pixels, 
with the pixel size 
being determined by the beam size. 

We then introduced the detector effects, first by convolving the lensed map with the beam profile 
$b_{\ell}=\sum_{i} \exp[-({\theta_{fwhm}}_{i})^2\ell(\ell+1)/(8\ln2)/2]$
and then by adding the map constructed from the detector noise power spectrum
$1/\sum_{i}[ 1/\left({\theta_{fwhm}}_{i}{\sigma_{pix}}_{i}\right)^2],$
where the sums are over the number of channels $i.$
Finally we deconvolved the resulting map with the beam profile.

\section{Implementation of the harmonic space estimator}
\label{app:harmonic_estimator_implementation}

In this Appendix we describe the implementation of this estimator to reconstruct the convergence field $\kappa_{0}.$
The optimal estimator is a convolution in harmonic space of the functions 
$F_1(\boldsymbol{\theta})$ and 
$\nabla_{\boldsymbol{\theta}} F_{2}(\boldsymbol{\theta}),$ 
which are defined as follows
\ba
F_{1}(\boldsymbol{\ell}) 
&=&{1\over {\tilde C_{\ell}+N_{\ell} }}
  ~\tilde T(\boldsymbol{\ell}) \\
F_{2}(\boldsymbol{\ell})
&=&{C_{\ell}\over 
{\tilde C_{\ell}+N_{\ell} }}
  ~\tilde T(\boldsymbol{\ell}),
\ea
so that
\ba
\hat \psi(\boldsymbol{\ell})
&=&-i~\mathcal{N}_{\ell}
\int {d^2\boldsymbol{\ell}^{\prime}\over (2\pi)^2}~
\boldsymbol{\ell}\cdot\boldsymbol{\ell}^{\prime}~
 F_{1}(\boldsymbol{\ell}-\boldsymbol{\ell}^{\prime})~
 (\nabla_{\boldsymbol{\theta}}F_{2})(\boldsymbol{\ell}^{\prime})\cr
&=&-i~\mathcal{N}_{\ell}~\boldsymbol{\ell}\cdot
\int d^2\boldsymbol{\theta}~
\exp[-i\boldsymbol{\ell}\cdot\boldsymbol{\theta}]~
 F_{1}(\boldsymbol{\theta})~
 \nabla_{\boldsymbol{\theta}}F_{2}(\boldsymbol{\theta})\cr
&=&-\mathcal{N}_{\ell}
\int d^2\boldsymbol{\theta}~
\exp[-i\boldsymbol{\ell}\cdot\boldsymbol{\theta}]~
 \nabla_{\boldsymbol{\theta}}\left[
  F_{1}(\boldsymbol{\theta})~
  \nabla_{\boldsymbol{\theta}}F_{2}(\boldsymbol{\theta}) \right].
\ea
The estimator of the lensing field measures the correlations between the small-scale filtered temperature, as encoded in $F_{1}(\boldsymbol{\theta}),$ and the temperature gradient, as encoded in 
$\nabla_{\boldsymbol{\theta}}F_2(\boldsymbol{\theta}).$
This result can be intuited by noting that
while on small scales the unlensed temperature anisotropy is approximately a temperature gradient,
the small-scale lensed anisotropy comes from disturbing this gradient by the deflection angle from the lensing potential.
Note that the quadratic estimator is only valid to lowest order in the lensing potential.

From the lensed CMB map previously synthesised, 
we first construct the two scalar fields $F_1(\boldsymbol{\ell})$ and $F_{2}(\boldsymbol{\ell}).$
We then construct
two high-pass filters in real space, one explicitly from $F_{1}(\boldsymbol{\ell})$ and the other from the gradient of $F_{2}(\boldsymbol{\ell})$
\ba
F_1(\boldsymbol{\theta})
&=&
\int {d^2\boldsymbol{\ell}\over {(2\pi)^2}}~
 F_{1}(\boldsymbol{\ell})~
 e^{i\boldsymbol{\ell}\cdot \boldsymbol{\theta}}
\\
\nabla_{\boldsymbol{\theta}}F_2(\boldsymbol{\theta})
&=&
\int {d^2\boldsymbol{\ell}\over {(2\pi)^2}}~i\boldsymbol{\ell}~
 F_{2}(\boldsymbol{\ell})~
 e^{i\boldsymbol{\ell}\cdot \boldsymbol{\theta}}.
\ea
We can regard the first filter as a weight of the gradient of the temperature map (the second filter), resulting the vector field $\boldsymbol{F}_{3}(\boldsymbol{\theta})$ 
\ba
\boldsymbol{F}_{3}(\boldsymbol{\theta})
=F_{1}(\boldsymbol{\theta})~
\nabla_{\boldsymbol{\theta}}F_{2}(\boldsymbol{\theta}).
\ea
We then take the divergence of $\boldsymbol{F}_{3}(\boldsymbol{\theta})$ and transform into the $\boldsymbol{\ell}$ space, defining the scalar field 
$F_{4}(\boldsymbol{\ell})$
\ba
F_{4}(\boldsymbol{\ell})
=i\boldsymbol{\ell}\cdot \boldsymbol{F}_{3}(\boldsymbol{\ell}).
\ea 
This quantity is essentially the quadratic estimator of the gravitational potential $\psi$ 
up to the total variance $\cal{N}_{\ell}$
\ba
\hat \psi (\boldsymbol{\ell})
=-{\cal N}_{\ell}~F_{4}(\boldsymbol{\ell}).
\ea
(See Appendix~\ref{app:total_variance} for the parametrization adopted to calculate 
$\cal{N}_{\ell}.$) 
Finally we compute
\ba
\hat {\boldsymbol \alpha}(\boldsymbol{\ell})
&=&i\boldsymbol{\ell}~\hat \psi(\boldsymbol{\ell})
={\cal N}_{\ell}~\boldsymbol{\ell}~
[\boldsymbol{\ell}\cdot\boldsymbol{F}_{3}(\boldsymbol{\ell})]\\
\hat \kappa_{0}(\boldsymbol{\ell})
&=&
\ell ^2~\hat \psi(\boldsymbol{\ell}).
\ea

\section{Calculation of the total variance}
\label{app:total_variance}

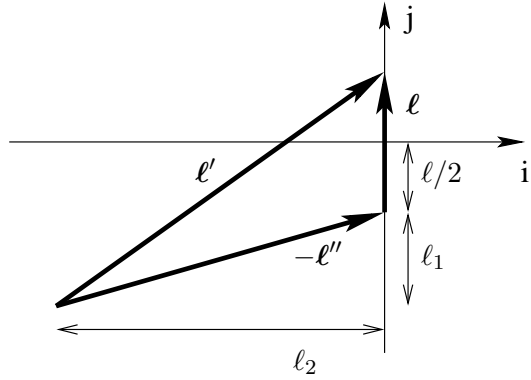
\begin{figure}
\setlength{\unitlength}{1cm}
\includegraphics[width=7cm]{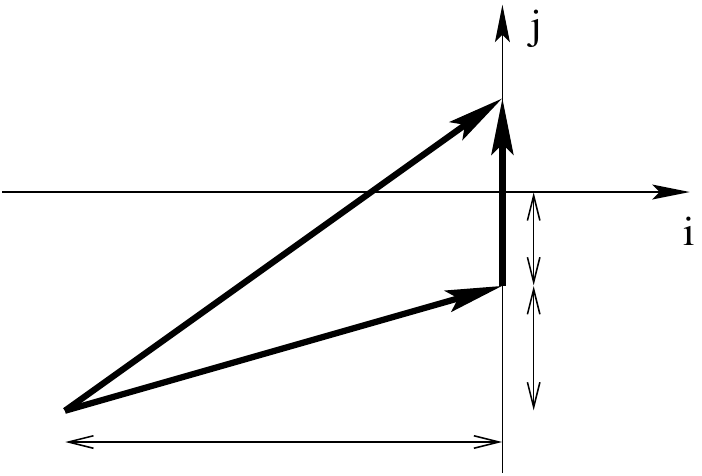}
\put(-1.7,3.2){$\boldsymbol{\ell}$}
\put(-4.5,2.3){$\boldsymbol{\ell}^{\prime}$}
\put(-3.2,1.2){$-\boldsymbol{\ell}^{\prime\prime}$}
\put(-1.5,2.3){$\ell/2$}
\put(-1.5,1.2){$\ell_{1}$}
\put(-3.2,-0.2){$\ell_{2}$}
\caption{\baselineskip=0.5cm{
{\bf Geometric relation of the position vectors in harmonic space.}
}}
\label{fig:pikassoyla_nl}

\end{figure}

In this Appendix we detail the coding of the integrand function for the calculation of  the total variance $\cal{N}_{\ell}$ given by Eqn.~(\ref{eqn:norm_factor}). First we demonstrate the calculation in the parametrization $(\boldsymbol{\ell},\boldsymbol{\ell}^{\prime})$ and then extend the reasoning to the parametrization $(\boldsymbol{\ell}_{+},\boldsymbol{\ell}_{-}).$

Since the intervening vectors form a triangle, we adopt for coordinate system an orthogonal grid $(\boldsymbol{i},\boldsymbol{j})$ such that the $\boldsymbol{j}$ axis coincides with the direction of $\boldsymbol{\ell},$ i.e. $\boldsymbol{\ell}=\ell\boldsymbol{j}.$ The other two vectors $\boldsymbol{\ell}^{\prime}$ and $\boldsymbol{\ell}^{\prime\prime}=\boldsymbol{\ell}-\boldsymbol{\ell}^{\prime}$ have components along these axes as follows  (see Fig.~\ref{fig:pikassoyla_nl})
\ba
\boldsymbol{\ell}^{\prime}
&=&\ell_{2}\boldsymbol{i}+\left(\ell_{1}+{\ell\over 2}\right)\boldsymbol{j}\\
\boldsymbol{\ell}^{\prime\prime}
&=&-\ell_{2}\boldsymbol{i}-\left(\ell_{1}-{\ell\over 2}\right)\boldsymbol{j}.
\ea
In this coordinate system we work out the quantities 
\ba
\boldsymbol{\ell}\cdot\boldsymbol{\ell}^{\prime}
=\ell\ell_{1}+{\ell^2\over 2}, \quad
\boldsymbol{\ell}\cdot\boldsymbol{\ell}^{\prime\prime}
=-\ell\ell_{1}+{\ell^2\over 2}
\ea
as well as 
\ba
{\boldsymbol{\ell}^{\prime}}^2=\ell_{2}^2+\left(\ell_{1}+{\ell\over 2}\right)^2,
\quad
{\boldsymbol{\ell}^{\prime\prime}}^2=\ell_{2}^2+\left(\ell_{1}-{\ell\over 2}\right)^2
\ea
and implement a two-dimensional numerical integration routine over $\ell_{1}$ and $\ell_{2},$ with $\ell$ as an input, to compute $\cal{N}_{\ell}$.

When we instead work in the coordinates $(\boldsymbol{\ell}_{+},\boldsymbol{\ell}_{-}),$ we note that $\boldsymbol{\ell}=\boldsymbol{\ell}_{+}$ and
$\boldsymbol{\ell}^{\prime}=(\boldsymbol{\ell}_{+}+\boldsymbol{\ell}_{-})/2$ which implies that
$\boldsymbol{\ell}^{\prime\prime}=(\boldsymbol{\ell}_{+}-\boldsymbol{\ell}_{-})/2.$ Following the same reasoning as above, we find the same component decomposition for $\boldsymbol{\ell}^{\prime}$ and $\boldsymbol{\ell}^{\prime\prime}$ along the $(\boldsymbol{i},\boldsymbol{j})$ axes, only with $\ell$ replaced by $\ell_{+}.$

\hfill

\end{document}